\documentclass[aps,prx,twocolumn,letterpaper,nobibnotes,superscriptaddress,longbibliography]{revtex4}
\usepackage{amssymb}
\usepackage{amsmath}
\usepackage{graphicx}
\usepackage{extarrows}
\usepackage{url}
\usepackage{varioref}
\usepackage{hyperref}
\usepackage{cleveref}
\usepackage{lineno}
\usepackage{color}
\usepackage{ulem}

\usepackage{float}
\usepackage{verbatim}

\usepackage{amstext}
\usepackage{epstopdf}
\usepackage{booktabs}
\usepackage{bm}
\usepackage{appendix}
\usepackage[utf8]{inputenc}
\usepackage{multirow}
\usepackage{bbold}
\usepackage{bbm}
\usepackage{latexsym}
\usepackage{xcolor}
\usepackage{braket}
\usepackage{wasysym}
\usepackage{verbatim}

\definecolor{lblue} {RGB}{51,71,158}

\begin{document}

\title{Probing Quantum Many-Body Correlations by Universal Ramping Dynamics}

\author{Libo Liang}
\thanks{These authors contribute equally to this work.}
\affiliation{School of Electronics, Peking University, Beijing 100871, China}
\author{Wei Zheng}
\thanks{These authors contribute equally to this work.}
\affiliation{Hefei National Laboratory for Physical Sciences at the
Microscale and Department of Modern Physics, University of Science and Technology of China, Hefei 230026, China}
\affiliation{CAS Center for Excellence in Quantum Information and Quantum Physics, University of Science and Technology of China, Hefei 230026, China}
\author{Ruixiao Yao}
\thanks{These authors contribute equally to this work.}
\affiliation{Department of Physics and State Key Laboratory of Low Dimensional Quantum Physics, Tsinghua University, Beijing 100084, China}
\author{Qinpei Zheng}
\affiliation{School of Electronics, Peking University, Beijing 100871, China}
\author{Zhiyuan Yao}
\author{Tian-Gang Zhou}
\affiliation{Institute for Advanced Study, Tsinghua University, Beijing 100084, China}
\author{Qi Huang}
\affiliation{School of Electronics, Peking University, Beijing 100871, China}
\author{Zhongchi Zhang}
\author{Jilai Ye}
\affiliation{Department of Physics and State Key Laboratory of Low Dimensional Quantum Physics, Tsinghua University, Beijing 100084, China}
\author{Xiaoji Zhou}
\affiliation{School of Electronics, Peking University, Beijing 100871, China}
\author{Xuzong Chen}
\email{xuzongchen@pku.edu.cn}
\affiliation{School of Electronics, Peking University, Beijing 100871, China}
\author{Wenlan Chen}
\email{cwlaser@ultracold.cn}
\affiliation{Department of Physics and State Key Laboratory of Low Dimensional Quantum Physics, Tsinghua University, Beijing 100084, China}
\affiliation{Frontier Science Center for Quantum Information, Beijing, 100084, China}
\author{Hui Zhai}
\email{hzhai@mail.tsinghua.edu.cn}
\affiliation{Institute for Advanced Study, Tsinghua University, Beijing 100084, China}
\author{Jiazhong Hu}
\email{hujiazhong01@ultracold.cn}
\affiliation{Department of Physics and State Key Laboratory of Low Dimensional Quantum Physics, Tsinghua University, Beijing 100084, China}
\affiliation{Frontier Science Center for Quantum Information, Beijing 100084, China}

\begin{abstract}
Ramping a physical parameter is one of the most common experimental protocols in studying a quantum system, and ramping dynamics has been widely used in preparing a quantum state and probing physical properties. Here, we present a novel method of probing quantum many-body correlation by ramping dynamics. We ramp a Hamiltonian parameter to the same target value from different initial values and with different velocities, and we show that the first-order correction on the finite ramping velocity is universal and path-independent, revealing a novel quantum many-body correlation function of the equilibrium phases at the target values. We term this method as the non-adiabatic linear response since this is the leading order correction beyond the adiabatic limit. We demonstrate this method experimentally by studying the Bose-Hubbard model with ultracold atoms in three-dimensional optical lattices. Unlike the conventional linear response that reveals whether the quasi-particle dispersion of a quantum phase is gapped or gapless, this probe is more sensitive to whether the quasi-particle lifetime is long enough such that the quantum phase possesses a well-defined quasi-particle description. In the Bose-Hubbard model, this non-adiabatic linear response is significant in the quantum critical regime where well-defined quasi-particles are absent. And in contrast, this response is vanishingly small in both superfluid and Mott insulators which possess well-defined quasi-particles. Because our proposal uses the most common experimental protocol, we envision that our method can find broad applications in probing various quantum systems.
\\
\textbf{keywords: }ramping dynamics, many-body correlations, optical lattices, degenerate quantum gas
\end{abstract}
%\jelcode{Pa, J6, P16, E22}

\maketitle

\section{Introduction}

Quantum many-body systems display rich phenomena characterized by varieties of correlations,
and many experimental tools have been developed to probe these correlations.
These methods include various spectroscopies and transport measurements in both condensed matter
systems \cite{RevModPhys.93.025006,neutron_scatterings_Rev@2019,transport@2005} and ultracold atomic systems \cite{Stringari,Zhai}. These probes can measure quasi-particle dispersions and reveal whether a quantum phase possesses a charge gap or spin gap, with the help of the linear response theory.
For instance, possessing a gap or not is an important way to characterize quantum many-body correlations and distinguishes different phases.

%For instance, metals and Bose superfluids are gapless, whereas band or Mott insulators display charge gap, and $s$-wave fermion paired superfluids exhibit spin gap.

There is also another important aspect of quantum many-body correlations, that is, whether the quasi-particle lifetime is long enough such that a quantum phase possesses a well-defined quasi-particle description or not \cite{Sachdev2011}.
It is a different characterization of quantum phases, compared with gap or gapless feature in the dispersion.
Quantum phases, such as conventional metals, band insulators, and Bose or Fermi superfluids, have well-defined quasi-particles. Among them, some are gapless, such as metals and Bose superfluids. And some
are gapped, for instance, $s$-wave fermion paired superfluids have spin gaps and band insulators have charge gaps.
Quantum phases, such as states in quantum critical regimes \cite{Sachdev2011}, Luttinger liquids in one-dimension \cite{giamarchi2003quantum} and non-Fermi
liquids \cite{nonFL@2001.Stewart,Lee2018}, do not have well-defined quasi-particle descriptions.

In both condensed matter and ultracold atomic systems, spectroscopy measurements can always determine the entire spectral function \cite{RevModPhys.93.025006,neutron_scatterings_Rev@2019,Stringari,Zhai,spectroscopy1,spectroscopy2,spectroscopy3,spectroscopy4,spectroscopy5,spectroscopy6,spectroscopy7}. Once the entire spectral function is known, it becomes clear whether a system is gapped or whether the system has a well-defined quasi-particle description. However, such measurements require scanning all frequency ranges in the relevant energy scale. For many properties, there is a more direct measurement that is less involved. A typical example is the charge gap. A dc transport experiment can immediately tell whether the system has a charge gap without knowing the complete information of the spectral function. This work will propose a similar shortcut to measure whether the system has well-defined quasi-particle behaviors, probing via ramping dynamics in many-body systems.

Ramping a physical parameter is one of the most widely-used control protocols in studying a quantum system. When the ramping rate is slow enough, the quantum state can follow the change of parameters adiabatically
and retain the ground state of the instantaneous Hamiltonian at a given
time. This protocol has been widely used in preparing a quantum state with
high fidelities and adiabatic quantum computations. When the ramping rate is non-negligible, the system is brought into a non-equilibrium situation that deviates from the instantaneous ground state and generates
excitations. In this situation, the ramping protocol can be turned into a
probing scheme, and two of the most well-known examples are the Thouless
pumping \cite{1983Quantization,Pump@1984.Niu,2016Topological,2016A} and the Kibble-Zurek mechanism \cite{Kibble_1976,Kibble_1980,Zurek1985Cosmological,2005Dynamics,2011Quantum,2014Emergence,2016Universal,2016Quantum,2019Quantum,2019Shin,Pan2019}.
For the Thouless pumping, the accumulated charge is quantized after a
pumping cycle, and this quantized charge probes the topological invariant
of the equilibrium phase \cite{1983Quantization,Pump@1984.Niu,2016Topological,2016A}. For the Kibble-Zurek mechanism, topological defects are excited when a
parameter is ramped across an equilibrium phase transition point, and the
dependence of topological defect numbers on ramping rates reveals the
critical exponent of the equilibrium phase transition \cite{Kibble_1976,Kibble_1980,Zurek1985Cosmological,2005Dynamics,2011Quantum,2014Emergence,2016Universal,2016Quantum,2019Quantum,2019Shin,Pan2019}.

Here we present a novel scheme of probing quantum many-body
correlations by ramping dynamics, with both theoretical frameworks and
experimental results. Our scheme utilizes the first-order correction on finite ramping rates beyond the adiabatic limit, and therefore, we term it as \textit{the non-adiabatic linear response}. Remarkably, we show that the
response is independent of the history of the ramping trajectories and only depends on the ending point of the ramping. In other words, our scheme probes the universal aspects instead of the details of the ramping dynamics. Moreover, the universal quantity deduced from this response can be attributed to an equilibrium quantum many-body correlation function at the ending point. Unlike the Thouless pumping and the Kibble-Zurek mechanism, the correlation function revealed by this method is quite general, not limited to topology or criticality. We investigate this scheme numerically in three different models as examples, the transverse field Ising model, the fermion pairing model, and the Bogoliubov model for bosons. We also demonstrate this scheme experimentally by studying the Bose-Hubbard model using degenerate bosonic atoms in optical lattices. In the Bose-Hubbard model, we show theoretically and experimentally that this response is significant in the quantum critical regime without well-defined quasi-particles and is vanishingly small in the superfluid phase (gapless) and the bosonic Mott insulator phase (gapped), both of which possess well-defined quasi-particle descriptions. Therefore, our results show that this response can be sensitive to whether the quantum phases possess well-defined quasi-particle descriptions rather than whether their quasi-particle dispersions are gapless or gapped.

\begin{figure}
\begin{center}
\includegraphics[width=0.49\textwidth]{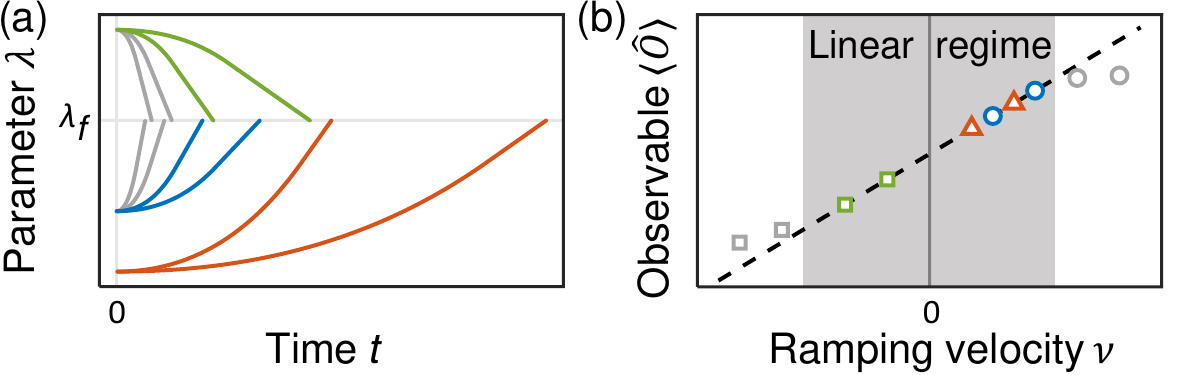}
\end{center}
\caption{\textbf{Schematic of the main result.} (a) Ramping a
parameter $\protect\lambda$ in Hamiltonian $H(\protect\lambda)$ to the same
final value $\protect\lambda_f$ with different initial values and different
ramping velocities $\protect\nu$. Measurement $\langle \hat{O}\rangle$ is
performed right after the ramping dynamics ends at $\protect\lambda_f$. (b) The measured $\langle \hat{O}\rangle$ with various ramping
trajectories in terms of ramping velocity $\protect\nu$. In the region of small $%
\protect\nu$ indicated by the shaded area, all data points collapse into a linear curve and the slope $%
\protect\alpha$ of the curve only depends on the final value $\protect\lambda%
_f$, independent of initial values and other details of the trajectories. This
slope probes the equilibrium correlation at $\protect\lambda_f$ given by Eq.
\protect\ref{alpha}. }
\label{ramping_illustration}
\end{figure}

\section{Theoretical Framework}
Let's consider a Hamiltonian $\hat{H}%
(\lambda)$ that depends on a parameter $\lambda$, and a time-dependent
ramping of the parameter $\lambda(t)$ from $\lambda_i$ to $\lambda_f$.
We start with the ground state at $\lambda_i$ and we choose $\lambda(t)$ that satisfies i) $
\partial\lambda/\partial t|_{\lambda_i}=0$; ii) $\partial\lambda/\partial
t|_{\lambda_f}=\nu$; and iii) the absolute value of $\partial\lambda/\partial t$ is always bounded
by $\nu$ for the entire ramping duration. As soon as $\lambda$ reaches $%
\lambda_f$, we immediately measure an observable $\langle\hat{O}\rangle$.
Suppose we repeat the measurements with different ramping trajectories, by choosing different initial state at different $\lambda_i$ and different ramping velocities, as shown in Fig. 1a, and then we plot $\langle\hat{O}\rangle$
as a function of the velocity $\nu$, as schematically shown in Fig. 1b. We can make a series expansion of $\langle%
\hat{O}\rangle$ in term of $\nu$ as
\begin{equation}
\langle\hat{O}\rangle=\langle\lambda_f|\hat{O}|\lambda_f\rangle+\alpha\nu+%
\dots  \label{expansion1}
\end{equation}
Here $|\lambda\rangle$ denotes the instantaneous ground state of $\hat{H}%
(\lambda)$ and $\nu$ can be either positive or negative. The leading term in Eq. \ref{expansion1} follows the
adiabatic approximation at $\nu\rightarrow 0$ and only depends on the
instantaneous ground state $|\lambda_f\rangle$ at the ending point of the
ramping.

Since $\langle\hat{O}\rangle$ in Eq. \ref{expansion1} is measured under the
instantaneous quantum state following the ramping dynamics, $\langle\hat{O}%
\rangle$ should depend on the entire ramping trajectory. However, the
main finding of this work is that, under the conditions (i)-(iii) mentioned above, the coefficient $\alpha$ of the
linear term in Eq. \ref{expansion1} only depends on the quantum state at the
ending point and is independent of the starting point $\lambda_i$, and other detail of the trajectory. That is to say, the results measured with
different ramping trajectories shown in Fig. 1a should collapse into a single straight line in the regime of
small $\nu$, and the slope of this line determines $\alpha$, as
schematically shown in Fig. 1b. Moreover,
we find that $\alpha$ measures the correlation function at the ending point
given by
\begin{equation}
\alpha=i\frac{\partial \mathcal{G}^R(\omega,\lambda_f)}{\partial \omega}\Big|%
_{\omega=0}.  \label{alpha}
\end{equation}
Here $\mathcal{G}^R(\omega,\lambda_f)$ is the Fourier transformation of the retarded Green's function $%
\mathcal{G}^R(t,\lambda_f)$, and $\mathcal{G}^R(t,\lambda_f)$ is defined as \cite{Sachdev2011}
\begin{equation}
\mathcal{G}^R(t,\lambda_f)=-i\Theta(t)\langle\lambda_f|[\hat{O}(t),\hat{V}%
(0)]|\lambda_f\rangle,  \label{GR}
\end{equation}
where $\hat{V}=\partial\hat{H}/\partial\lambda$ and $\Theta(t)$ is the step
function. In practice, this allows us to experimentally access the equilibrium correlation
given by Eq. \ref{alpha} by ramping to a given final parameter $\lambda_f$
with various ramping velocities. Since this correlation is obtained by the first
order correction away from the adiabatic limit, it is now termed as the
non-adiabatic linear response. Note that unlike the conventional linear response that is related to correlation functions, this response is related to the frequency derivative of correlation functions. As we will show below, this correlation function directly probes whether the spectral function is symmetric with respect to positive and negative frequencies and, therefore, provides direct access to the nature of quasi-particle description.

The proof of this result follows
straightforwardly from the perturbation expansion in term of ramping velocity, as we show in Supplementary Materials I.
In Supplementary Materials II, we also show three
examples, including the transverse field Ising model, the fermion pairing
model and the Bogoliubov model for bosons.
The numerical simulations of the ramping dynamics in these models confirm the consistency between the
slope and the correlation function given by Eq. \ref{alpha}. We remark that, although Eq. \ref{alpha}
and Eq. \ref{GR} are derived at zero-temperature, we can extend the formula
to finite temperature under the condition that the thermalization time scale is much
shorter than the ramping time scale. At finite temperature, we use the
thermal ensemble average to
replace the average over quantum state $|\lambda_f\rangle$ in Eq. \ref{GR}.

Here we should note that our theory is a perturbative expansion in terms of $\nu$. Therefore, there always exists a convergent regime where our theory is valid, as long as the linear order coefficient does not vanish and the higher order coefficients do not diverge, and this condition can be satisfied even for gapless systems. In the low dimension, the low-energy density-of-state is generically high, which leads to a high population of low-energy modes during the ramping dynamics. This leads to the divergence of high-order coefficients, consistent with the discussion of the breakdown of adiabaticity in low-energy gapless systems in the previous literature \cite{Polkovnikov2005,Polkovnikov2008}. We discuss the convergence conditions in more detail in Supplementary Materials III. As shown in Supplementary Materials III, if the ramping term and the observable both obey certain symmetry, the linear response will vanish due to the symmetry constraint. Hence, our discussion below always focuses on the cases without such symmetry. Under these conditions, we can always further expand Eq. \ref{expansion1} as
\begin{equation}
\langle\hat{O}\rangle=\langle\lambda_f|\hat{O}|\lambda_f\rangle+\alpha\nu+\beta \nu^2
+\dots,  \label{expansion-high}
\end{equation}
and the validity of the linear expansion at least requires $\nu\ll \alpha/\beta$. Note that $\beta$ is not a universal number and is path-dependent. Therefore, the validity range of the linear expansion is path-dependent.

We should also note the difference between our theory and the Kibble-Zurek mechanism. The Kibble-Zurek mechanism focuses on topological defects related to the long-range correlation of order parameters. Therefore, it experiences a critical slowing down at the critical point as it takes a long time to establish a long-range correlation \cite{Zurek1996,Zurek2014}. Whereas our theory only concerns local equilibrium, its validity is not affected by the critical slowing down. Hence, our theory can also be applied to ramping across a critical regime.

\begin{figure}
\begin{center}
\includegraphics[width=0.49\textwidth]{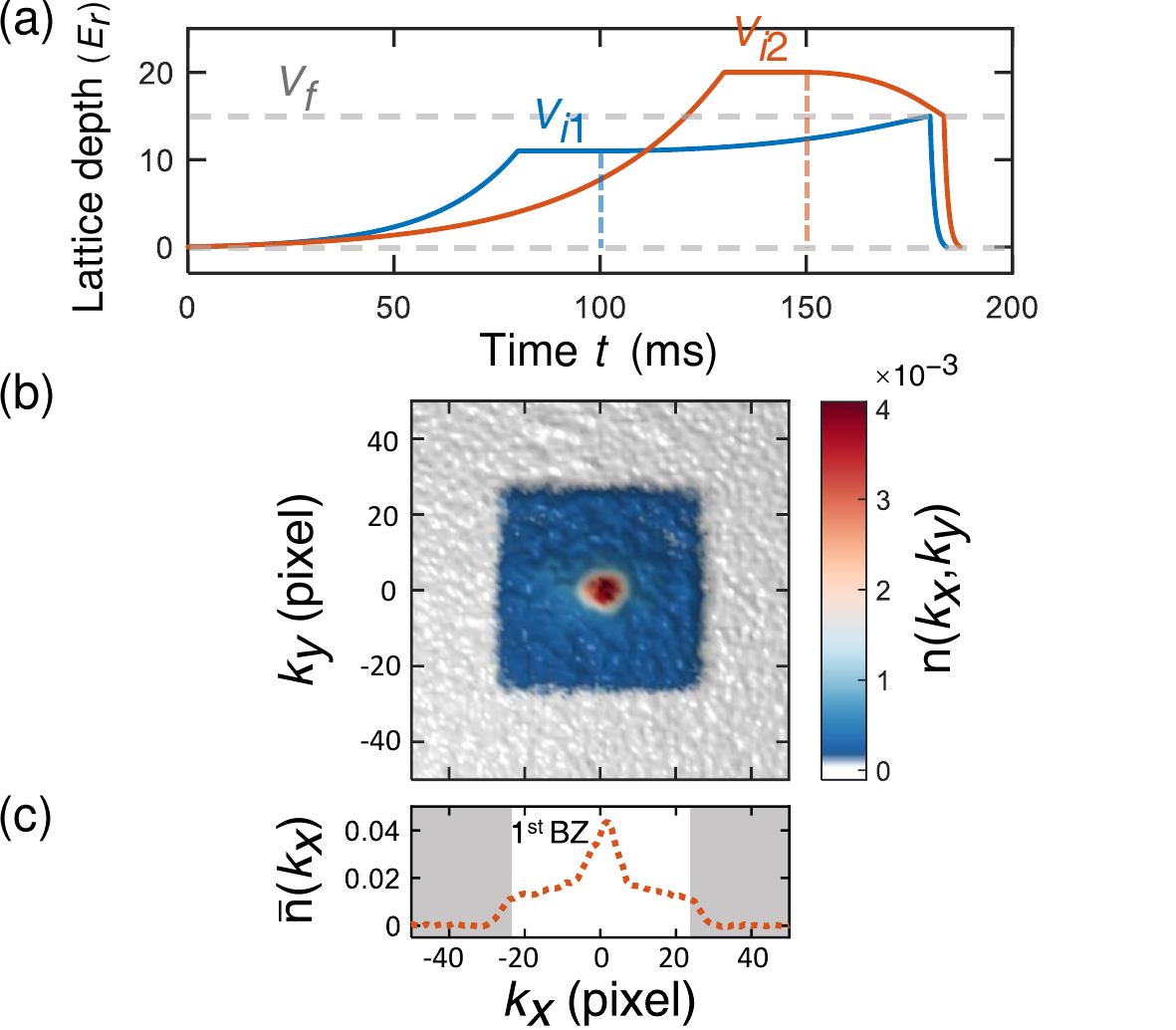}
\end{center}
\caption{\textbf{Experimental time sequence and typical results.} (a) The
time sequence of our experiments. We adiabatically load degenerate ${}^{87}$%
Rb into optical lattices with different initial lattice depths $V_i$, such
as $V_{i1}=11E_r$ (blue) and $V_{i2}=20E_r$ (red) in the illustration. The
atoms are hold at the initial lattice depth for $20$~ms, and then, we start to ramp the lattices after the
time indicated by the dashed lines. After a
smoothing procedure at the initial ramping process, we linearly ramp the lattice to the final depth $V_f$.
(b) Typical raw data of the band mapping measurement,
resulting in a two-dimensional quasi-momentum distribution $n(k_x,k_y)$.
(c) One-dimensional quasi-momentum distribution $\bar{n}%
(k_x)$ after integration over $k_y$. }
\label{exp_scheme}
\end{figure}

\section{Experimental Results}
The experiment in the Bose-Hubbard model is carried out with degenerate
${}^{87}$Rb atoms in a three-dimensional optical lattice.
The optical lattice is formed by three standing waves perpendicular to each other at wavelength $\lambda=1064$~nm and the magnetic field is applied along $z$ axis.
Each lattice beam has a beam waist of 150(10) $\mu$m while the atoms occupy a region with a radius of 13 $\mu$m.
When the lattice depth is at 5$E_r$ ($E_r=h\times 2$~kHz is the recoil energy of the optical lattice), the inhomogeneity of lattice beams provides an external harmonic trap with isotropic radial vibrational frequencies $\sim2\pi\times 20(1)$~Hz.
The ramping time sequence
of the experiment is shown in Fig. 2a. We
adiabatically load $1.6(1)\times10^5$ atoms into lattices with an initial lattice depth $V_i$
and hold the system for 20~ms for relaxations. Then we ramp the lattice
depth to $V_f$ with a velocity $\nu$ (in unit of $E_r/$ms). Here, the starting part
of the ramping curve is smoothened to satisfy conditions (i)-(iii) discussed above (see
Supplementary Materials IV for details). As soon as the lattice depth reaches $V_f$, we
perform the band-mapping measurement \cite{Kohl2005,huang2020observation} by imaging
the atoms along $z$-direction, and measure a two-dimensional quasi-momentum
distribution $n(k_x,k_y)$ of atoms. A typical result of the band mapping is
shown in Fig. 2b. We further integrate $n(k_x,k_y)$
along $k_y$-direction, which results in a one-dimensional quasi-momentum
distribution $\bar{n}(k_x)=\int dk_y n(k_x,k_y)$ as shown in Fig. 2c.

We ramp the lattice depth to the same target value $V_f=15E_r$ from
different initial lattice depths $V_i=5,11,17$ and $20E_r$, and measure $\bar{n}(k_x=0)$ as a function of $\nu$ for different $V_i$. We can see in Fig. 3a that there always exists a
linear regime and these linear regimes overlap with each other for
trajectories with different $V_i$. We extract the slope from the linear
regime and obtain the slope $\alpha$ of $0.025(2)$, $0.023(6)$, $%
0.024(4)$ and $0.025(3)$ for $V_i=5,11,17$ and $20E_r$ respectively as shown in Fig. 3b. We also get $\alpha$ of $0.025(2)$ and $0.024(2)$ for $
V_i=18$ and $19E_r$ from data shown in Fig. 4c.
Within the statistical errors, it is consistent with our theory that $\alpha$
is independent of the initial lattice depth $V_i$. Nevertheless, we should
note that for different $V_i$, the window of the linear regime is different.
This is because the higher order coefficients in the expansion Eq. \ref
{expansion1} depend on the initial value and other details of the
trajectories. As the higher order coefficients get larger, the linear
window gets smaller. We also note that, in the limit of $\nu\rightarrow 0$,
data taken with different $V_i$ should give the same result that recovers the
adiabatic limit. The small discrepancy in this limit between
different data sets (Fig. 4) is due
to the day-to-day drift of our experimental apparatus (see Supplementary Materials V).

We verify the path independence of $\alpha$ not only for $\bar{n}(k_x=0)$
but also for $\bar{n}(k_x)$ in the entire first Brillouin zone. Here, we
symmetrize the measured one-dimensional quasi-momentum distributions to extract $\bar{%
n}(k_x)$ in terms of $k_x$ (see Supplementary Materials VI). Fig. 3c and d show the slope $\alpha$ extracted from
$\bar{n}(k_x)$ as a function of $k_x$. Each plot shows results with the same
$V_f$ but two different $V_i$. One can see that, for the entire first
Brillouin zone, $\alpha(k_x)$ with the same $V_f$ and different $V_i$ coincide with
each other within the statistical errors.

Then, we vary $V_f$ to probe the correlations at different lattice depths. In
Fig. 4a$-$f, we show results for $V_f=11,13,15,17,19$ and $21E_r$. For each given $V_f$, we ramp the lattice
depth to this $V_f$ with at least two different $V_i$ and consistent slopes $\alpha$
are obtained for all cases. In Fig.~4g, we plot $\alpha$ as a function of $V_f$. We find that $\alpha$ is vanishingly
small for $V_f = 11E_r$ and $V_f = 21E_r$, and $\alpha$ is significant for $V_f$ in
the range between $13E_r$ and $19E_r$. Note that in our system, the
zero-temperature quantum phase transition between the superfluid and the Mott
insulator occurs at $13E_r$ for density $n=1$, $15E_r$ for $n=2$, and $%
17E_r$ for $n=3$ (the local density of our system is up to $n=3$). Hence,
the lattice depth $13\sim 19E_r$ corresponds to the quantum critical regime in
our system.

\begin{figure}
\begin{center}
\includegraphics[width=0.48\textwidth]{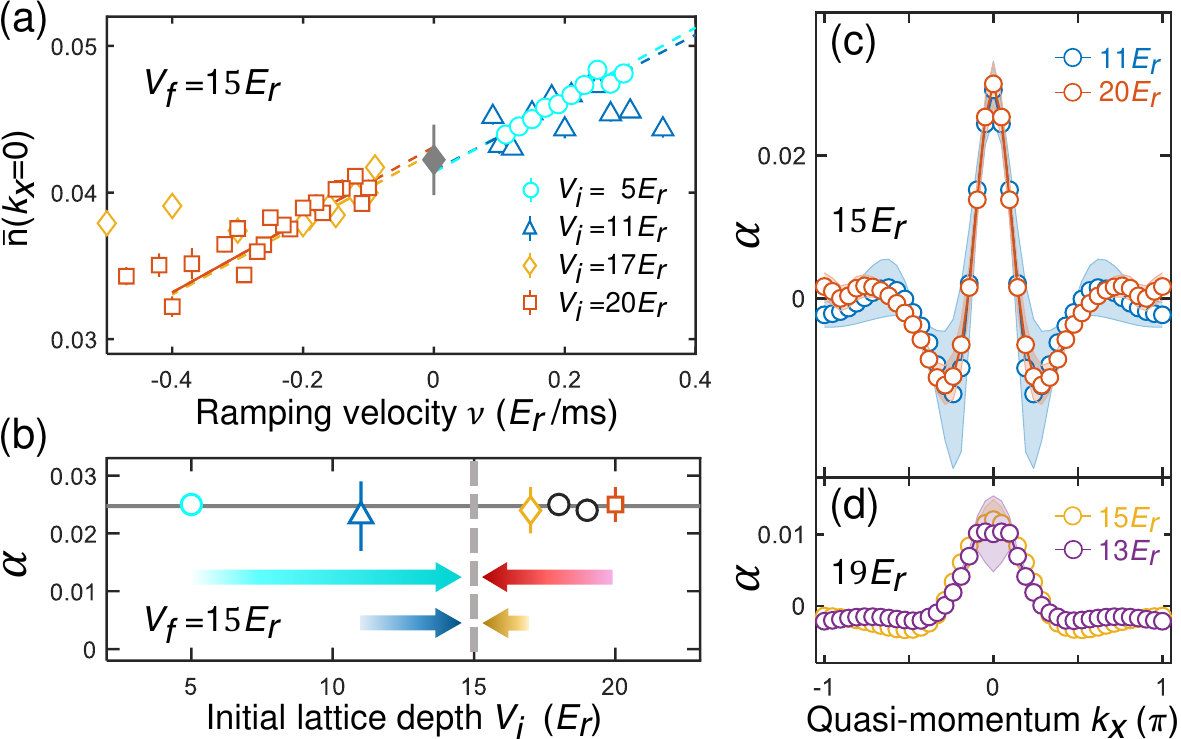}
\end{center}
\caption{\textbf{Experimental demonstration of the path independence.}
(a) $\bar{n}(k_x=0)$ versus the ramping velocity $\protect\nu$. Here we plot four
sets of data. The final lattice depth is fixed at $V_f=15E_r$, and the initial lattice depths are respectively $5E_r$ (cyan circle), $11E_r$ (blue triangle), $17E_r$ (yellow diamond) and $20E_r$ (red
square). The error bars here represent one standard errors of the
mean by repeating $20$ to $80$ measurements for each data point. The solid
lines are weighted linear fits to the data. The lengths of the solid lines
represent the fitting regime and the dashed lines are the extensions of the
linear fits. The cyan, blue, yellow, and red lines respectively yield slopes $\protect%
\alpha$ as $0.025(2)$, $0.023(6)$, $0.024(4)$, and $0.025(3)$. The grey diamond
labels the value of $\bar{n}(k_x=0)$ by adiabatically ramping to $V_f=15E_r$
whose error bar denotes one standard deviation of $386$ repeating
measurements. (b) $\protect\alpha$ versus the initial lattice
depth $V_i$. The horizontal solid line marks the mean value $0.025(1)$ of $
\protect\alpha$ which is obtained by the weighted average of $\protect
\alpha$ from six different $V_i$ with $V_i=5, 11, 17, 18, 19$, and $20E_r$. (c-d) $
\protect\alpha$ versus quasi-momentum $k_x$ for the entire first Brillouin
zone with $V_f=15E_r$ (\textbf{c}) and $V_f=19E_r$ (\textbf{d}). In \textbf{c%
}, blue circles represent the situation with $V_i=11E_r$ and the red circles represent the situation with $V_i=20E_r$. In \textbf{d}, yellow circles represent the situation with $V_i=15E_r$ and the purple circles represent the situation with
$V_i=13E_r$. The shadow areas denote
the range of one standard deviation due to statistical errors. The solid lines are
guides for eyes. }
\label{trajectory}
\end{figure}

Therefore, the experimental measurements not only confirm that the non-adiabatic
linear response is independent of the details of the ramping trajectories,
but also discover that this response is much more significant in the quantum critical
regime than that in the superfluid and the Mott insulator phases. To understand this result, we analyze the correlation function
probed by Eq. \ref{alpha} in the Bose-Hubbard model (BHM) below.

\begin{figure}
\begin{center}
\includegraphics[width=0.48\textwidth]{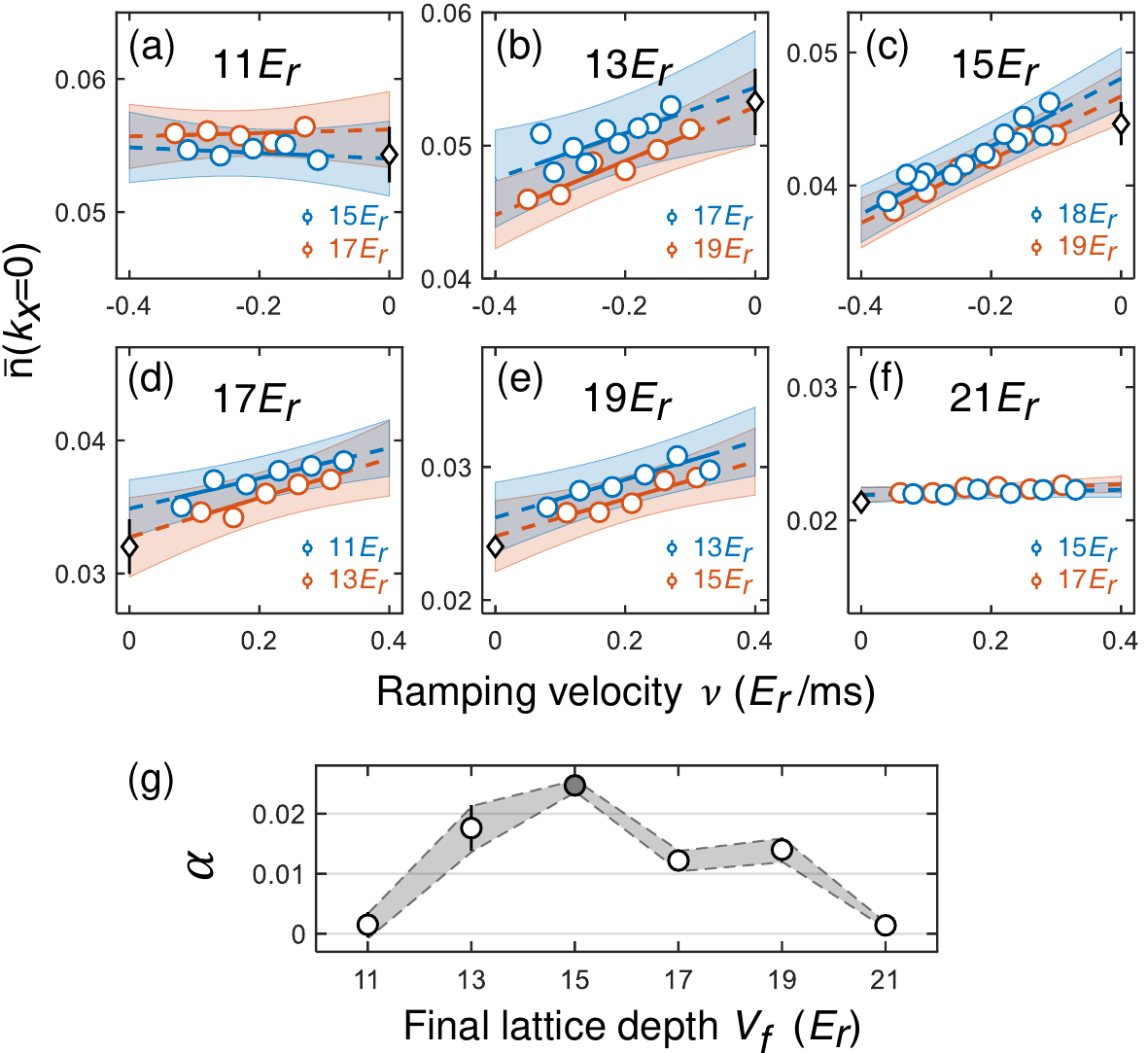}
\end{center}
\caption{ \textbf{The measured correlation versus the final
lattice depth $V_f$.} (a-f) $\bar{n}(k_x=0)$ versus the ramping velocity $%
\protect\nu$ by ramping to a set of different $V_f$ ($11, 13, 15, 17, 19$, or $21E_r$%
). Each panel show a fixed $V_f$ with two different initial $V_i$. The circles are
data with error bars (usually smaller than marker size) being one standard error of $20$ to $40$ repeated measurements. The solid lines are
the linear fits, and the dashed lines are the extension of linear fits
outside of the measurement ranges. The black diamonds correspond to the
adiabatic measurement of $\bar{n}(k_x=0)$ whose error bars are given by one standard
deviation of $15$ to $20$ repeated measurements. The shadow areas denote the 95\% confidence intervals. (g) $\alpha$ versus the final lattice depth $V_f$. The shadow area shows the
uncertainty range of one standard deviation. }
\label{critical}
\end{figure}

\section{Application to the Bose-Hubbard Model}
The Hamiltonian for the BHM is written as
\begin{eqnarray}
\hat{H}_{\text{BHM}}=& &-J \sum\limits_{\langle ij\rangle }(\hat{a}_{i}^{\dag }\hat{a}%
_{j}+h.c.)\nonumber \\
& &+\sum\limits_{i}\left[ \frac{U}{2}\hat{n}_{i}(\hat{n}_{i}-1)-\mu
\hat{n}_{i}\right] ,
\end{eqnarray}%
where $\hat{a}_{i}$ is the annihilation operator at site-$i$, $\hat{n}_{i}=%
\hat{a}_{i}^{\dag }\hat{a}_{i}$ is the particle number operator at site-$i$,
$J$ is the hopping strength between neighboring sites, and $U$ is the
on-site interaction strength. In the experiment, both $J$ and $U$ change in
time during ramping lattice depth. However, since the quasi-momentum
distribution is measured in experiments and the measurement operator $\hat{O}%
=\hat{n}_{\mathbf{k}}=\hat{a}_{\mathbf{k}}^{\dag }\hat{a}_{\mathbf{k}}$
commutes with the hopping term, the dominate effect during ramping should
come from the changing of parameter $U$. Hence, for simplicity, we consider ramping the interaction strength $U$ from an initial value $
U_{i}$ to a final value $U_{f}$, such that $\partial \hat{H}/\partial
\lambda=\sum_{i}\frac{1}{2}\hat{n}_{i}(\hat{n}_{i}-1)$. Note that the interaction
term can also be written in momentum space as
\begin{equation}
\frac{U}{2N_{s}}\sum\limits_{\mathbf{p},\mathbf{k_{1}},\mathbf{k_{2}}}\hat{a}%
_{\mathbf{p+k_{1}}}^{\dag }\hat{a}_{\mathbf{p-k_{1}}}^{\dag }\hat{a}_{%
\mathbf{p-k_{2}}}\hat{a}_{\mathbf{p+k_{2}}},
\end{equation}%
where $N_{s}$ is total number of sites. Thus, the non-adiabatic linear
response theory presented above probes the correlator
\begin{eqnarray}
\mathcal{G}^{R}(t,U_{f})&=&\frac{-i\Theta (t)}{2N_{s}}\sum\limits_{\mathbf{p},%
\mathbf{k_{1}},\mathbf{k_{2}}}\Big\langle\lbrack \hat{a}_{\mathbf{k}}^{\dag
}(t)\hat{a}_{\mathbf{k}}(t),\hat{a}_{\mathbf{p+k_{1}}}^{\dag }(0)\nonumber \\
& &\hat{a}_{%
\mathbf{p-k_{1}}}^{\dag }(0)\hat{a}_{\mathbf{p-k_{2}}}(0)\hat{a}_{\mathbf{%
p+k_{2}}}(0)]\Big\rangle.  \label{corre}
\end{eqnarray}
This correlator is different from density-density or phase correlation measured in the Bose-Hubbard model before \cite{density-density,phase}.

We implement the Wick decomposition to express the multiple-points
correlation function Eq. \ref{corre} in term of two-point correlation
functions, where the single-particle spectral function $%
\mathcal{A}(\mathbf{k},\omega)$ can be introduced through the two-point correlation
functions as \begin{align}
\Big\langle\hat{a}_{\mathbf{k}}^\dag(t)\hat{a}_{\mathbf{k^\prime}}(0)%
\Big\rangle=&\delta_{\mathbf{k},\mathbf{k^\prime}}\int d\omega f_B(\omega)%
\mathcal{A}(\mathbf{k},\omega)e^{i\omega t}, \\
\Big\langle\hat{a}_{\mathbf{k}}(t)\hat{a}_{\mathbf{k^\prime}}^\dag(0)%
\Big\rangle=&\delta_{\mathbf{k},\mathbf{k^\prime}}\int d\omega (1+f_B(\omega))%
\nonumber \\
&\times\mathcal{A}(\mathbf{k},\omega)e^{-i\omega t},
\end{align}
and $f_B(\omega)=1/(e^{\beta(\omega-\mu)}-1)$ is the Bose distribution
function (see Supplementary Materials VII and VIII). With this approximation, the correlator Eq. \ref{corre}, and
consequently $\alpha$ given by Eq. \ref{alpha}, is eventually determined by the
spectral function $\mathcal{A}(\mathbf{k},\omega)$ as
\begin{eqnarray}
\alpha =4\pi \bar{n}\int d\omega f_{\mathrm{B}}(\omega )\mathcal{A}\left( \mathbf{k,}\omega
\right) \frac{\partial}{\partial \omega} \mathcal{A}\left( \mathbf{k,}\omega \right) . \label{alpha-A}
\end{eqnarray}

In the BHM, there are two types of spectral function $\mathcal{A}(\mathbf{k}%
,\omega )$ \cite{Sachdev2011}. When the system is either deeply in the superfluid
phase or deeply in the Mott insulator phase, the system possesses well-defined
quasi-particles. In the case, $\mathcal{A}(\mathbf{k},\omega )$
behaves as
\begin{equation}
\mathcal{A}(\mathbf{k},\omega)\sim \frac{\Gamma_{{\bf k}}}{(\omega-\epsilon_{\mathbf{k}})^2+\Gamma^2_{{\bf k}}} ,
\end{equation}%
where $\epsilon _{\mathbf{k}}$ is the quasi-particle energy, and $1/\Gamma_{{\bf k}}$ gives the quasi-particle lifetime. When the quasi-particle lifetime is long enough, $\Gamma_{{\bf k}}\rightarrow 0$ and $k_{b}T \gg \Gamma_{{\bf k}}$. Then, $f_{\mathrm{B}}(\omega )$ can be taken as a constant in the energy window $\sim \Gamma_{{\bf k}}$ around $\epsilon_{{\bf k}}$. Thus, it is easy to see that $A\left( \mathbf{k,}\omega \right)$ is an even function and $\partial{A\left( \mathbf{k,}\omega \right)}/\partial \omega$ is an odd function centered around $\epsilon_{{\bf k}}$. Hence, after the integration, $\alpha$ approaches zero. When the system is in the critical regime, the system no longer possesses
well-defined quasi-particles and $\mathcal{A}(\mathbf{k},\omega )$
behaves as
\begin{equation}
\mathcal{A}(\mathbf{k},\omega )\sim \frac{\Theta (\omega -\epsilon _{\mathbf{%
k}})}{(\omega -\epsilon _{\mathbf{k}})^{\eta }},
\label{Acritical}
\end{equation}%
where $\eta $ is a critical exponent \cite{Sachdev2011,Higgs}. Substituting Eq. \ref{Acritical} into Eq. \ref{alpha-A}, it is straightforward to obtain
\begin{equation}
\alpha \propto \frac{n}{T^{2\eta }}f_{B}(\epsilon _{\mathbf{k}}).
\label{alpha_2}
\end{equation}%
This discussion explains the experimental findings presented in Fig. 4, and attributes the difference in the non-adiabatic linear
response in this system to whether the quantum phases possess well-defined
quasi-particle descriptions or not.

Ideally, by comparing our measurements with Eq. \ref%
{alpha_2}, we can determine the critical exponent by studying the
temperature dependence of this correlation. However, since our current
experiment is performed in the presence of a harmonic trap, the correlation
is smeared out by the density inhomogeneity in the real space. This
limitation can be lifted by using the box potential in a future experiment.

\section{Conclusions and outlook}
We find a new regime for many-body dynamics, where the deviations from steady states are independent of the trajectories of dynamics.
In this regime, the non-adiabatic response is linear instead of conventional power laws.
This provides us with a scheme to probe the many-body systems via universal ramping dynamics, and measure whether the system has well-defined quasi-particle behaviors.
Besides the BHM, our scheme can be directly applied to probe correlations in
other systems with ultracold atomic gases, such as unitary Fermi gas and
quantum simulation of various spin models. Our method can also be applied to
other systems beyond ultracold atomic gases, such as trapped ions, NV
centers, and condensed matter systems. As demonstrated in studying
the Bose-Hubbard model, our method accesses a different aspect of quantum
many-body correlation compared with many existing measurement tools. Thus,
our protocol provides a new tool for experimentally studying correlations in quantum
matters.

\section*{Acknowledgements}
This work was supported by Beijing Outstanding Young Scholar Program,
the National Key Research and Development Program of China (2021YFA0718303, 2021YFA1400904, and 2016YFA0301501),
 the National Natural Science Foundation of China (91736208, 11974202, 61975092, 11920101004, 61727819, 11934002, 11734010, and 92165203) and the XPLORER Prize.

\bibliographystyle{NSR_Yao}

\newpage
\setcounter{equation}{0}
\setcounter{figure}{0}
\setcounter{table}{0}
\setcounter{section}{0}
\renewcommand{\theequation}{S\arabic{equation}}
\renewcommand{\thefigure}{S\arabic{figure}}
\renewcommand{\thetable}{S\arabic{table}}
\large
\begin{center}
\textbf{Supplementary Materials}

\end{center}

\normalsize

\section{Derivation of the Non-Adiabatic Linear Response\label{appendA}}

We consider a time-dependent Hamiltonian $\hat{H}(\lambda(t))$ through
parameter $\lambda$ and time dependently ramp the parameter $\lambda (t)$
from $\lambda_{i}$ to $\lambda_{f}$. The instantaneous eigenstates and
eigenvalues of the Hamiltonian $\hat{H}(\lambda )$ are denoted as $
\left\vert \varphi _{m}(\lambda)\right\rangle $ and $E_{m}(\lambda )$, and
the instantaneous ground state is denoted as $\left\vert
\varphi_{0}(\lambda)\right\rangle $.

We start with the time-dependent Schr\"{o}dinger equation
\begin{equation}
i\partial _{t}\left\vert \psi (t)\right\rangle =H\left( \lambda \left(
t\right) \right) \left\vert \psi (t)\right\rangle,  \label{SGEQ}
\end{equation}
and we expand the wave function $\left\vert \psi (t)\right\rangle$ in term
of the instantaneous eigenstates $\left\vert \varphi _{m}(\lambda \left(
t\right) )\right\rangle $ as
\begin{equation}
\left\vert \psi (t)\right\rangle =\sum_{m}a_{m}\left( t\right) e^{-i\theta
_{m}\left( \lambda \left( t\right) \right) }\left\vert \varphi _{m}(\lambda
\left( t\right) )\right\rangle ,  \label{expansion}
\end{equation}%
where $a_{m}\left( t\right) $ is a time dependent coefficient, and
\begin{equation}
\theta_{m}\left( \lambda \right) =\int\nolimits_{\lambda _{i}}^{\lambda
}d\lambda ^{\prime }\left[ \frac{E_{m}(\lambda ^{\prime })}{\frac{\partial \lambda^{\prime}%
}{\partial t}}-\varepsilon _{m}\left( \lambda ^{\prime
}\right) \right],
\end{equation}
which is a phase factor with
\begin{equation}
\varepsilon _{m}\left( \lambda ^{\prime }\right) =\left\langle \varphi
_{m}(\lambda ^{\prime })\right\vert i\partial _{\lambda ^{\prime
}}\left\vert \varphi _{m}(\lambda ^{\prime })\right\rangle.
\end{equation}
Substituting Eq. \ref{expansion} into Eq. \ref{SGEQ}, we obtain%
\begin{align}
&i\partial _{t}a_{m}\left( t\right) =\nonumber\\
&-\frac{\partial \lambda }{\partial t}%
\sum_{n\neq m}\left\langle \varphi _{m}(\lambda )\right\vert i\partial
_{\lambda }\left\vert \varphi _{n}(\lambda )\right\rangle e^{i\theta
_{m}\left( \lambda \right) -i\theta _{n}\left( \lambda \right) }a_{m}\left(
t\right) .  \label{EQofa}
\end{align}%
Considering the situation that the ramping velocity $\left\vert \partial
\lambda /\partial t\right\vert $ is slow enough throughout the entire
ramping dynamics, one can solve this equation perturbatively in terms of
ramping velocity by expanding the solution as
\begin{equation}
a_{m}\left( t\right) =a_{m}^{\left( 0\right) }\left( t\right) +\frac{%
\partial \lambda }{\partial t}a_{m}^{\left( 1\right) }\left( t\right)
+\cdots .  \label{exp2}
\end{equation}%
Substituting Eq. \ref{exp2} into Eq. \ref{EQofa}, one obtains
\begin{align}
&i\partial _{t}a_{m}^{\left( 0\right) }\left( t\right) =0, \\
&i\partial _{t}a_{m}^{\left( 1\right) }\left( t\right) =\nonumber\\
&-\sum_{n\neq
m}\left\langle \varphi _{m}(\lambda )\right\vert i\partial _{\lambda
}\left\vert \varphi _{n}(\lambda )\right\rangle e^{i\theta _{m}\left(
\lambda \right) -i\theta _{n}\left( \lambda \right) }a_{m}^{\left( 0\right)
}\left( t\right), \\
&\cdots.
\end{align}%
The solution of the zeroth order equation is a constant denoted by $%
a_{m}^{\left( 0\right) }\left( t\right) =a_{m}^{\left( 0\right) }\left(
t_{i}\right) $. Then one can obtain the first order correction as
\begin{widetext}
\begin{eqnarray}
a_{m}^{\left( 1\right) }\left( t\right) -a_{m}^{\left( 1\right) }\left(
t_{i}\right) &=&\sum_{n\neq m}\left\{ i\int_{t_{i}}^{t}dt^{\prime
}\left\langle \varphi _{m}(\lambda \left( t^{\prime }\right) )\right\vert
i\partial _{\lambda }\left\vert \varphi _{n}(\lambda \left( t^{\prime
}\right) )\right\rangle e^{i\theta _{m}\left( \lambda \left( t^{\prime
}\right) \right) -i\theta _{n}\left( \lambda \left( t^{\prime }\right)
\right) }\right\} a_{n}^{\left( 0\right) }\left( t_{i}\right) , \\
&=&\sum_{n\neq m}\left\{ i\int_{\lambda _{i}}^{\lambda }d\lambda ^{\prime }%
\frac{\partial t}{\partial \lambda ^{\prime }}\left\langle \varphi
_{m}(\lambda ^{\prime })\right\vert i\partial _{\lambda ^{\prime
}}\left\vert \varphi _{n}(\lambda ^{\prime })\right\rangle e^{i\theta
_{m}\left( \lambda ^{\prime }\right) -i\theta _{n}\left( \lambda ^{\prime
}\right) }\right\} a_{n}^{\left( 0\right) }\left( t_{i}\right).  \label{am1}
\end{eqnarray}
Eq. \ref{am1} can be evaluated further following the integration by parts
\begin{eqnarray}
&&i\int_{\lambda _{i}}^{\lambda }d\lambda ^{\prime }\frac{\partial t}{%
\partial \lambda ^{\prime }}\left\langle \varphi _{m}(\lambda ^{\prime
})\right\vert i\partial _{\lambda ^{\prime }}\left\vert \varphi _{n}(\lambda
^{\prime })\right\rangle e^{i\theta _{m}\left( \lambda ^{\prime }\right)
-i\theta _{n}\left( \lambda ^{\prime }\right) }  =\int_{\lambda _{i}}^{\lambda }\frac{\left\langle \varphi _{m}(\lambda
^{\prime })\right\vert i\partial _{\lambda ^{\prime }}\left\vert \varphi
_{n}(\lambda ^{\prime })\right\rangle de^{i\theta _{m}\left( \lambda
^{\prime }\right) -i\theta _{n}\left( \lambda ^{\prime }\right) }}{%
E_{m}(\lambda ^{\prime })-E_{n}(\lambda ^{\prime })-\dot{\lambda}^{\prime }%
\left[ \varepsilon _{m}\left( \lambda ^{\prime }\right) -\varepsilon
_{n}\left( \lambda ^{\prime }\right) \right] }  \notag \\
&=&\left. \frac{\left\langle \varphi _{m}(\lambda ^{\prime })\right\vert
i\partial _{\lambda ^{\prime }}\left\vert \varphi _{n}(\lambda ^{\prime
})\right\rangle e^{i\theta _{m}\left( \lambda ^{\prime }\right) -i\theta
_{n}\left( \lambda ^{\prime }\right) }}{E_{m}(\lambda ^{\prime
})-E_{n}(\lambda ^{\prime })-\dot{\lambda}^{\prime }\left[ \varepsilon
_{m}\left( \lambda ^{\prime }\right) -\varepsilon _{n}\left( \lambda
^{\prime }\right) \right] }\right\vert _{\lambda _{i}}^{\lambda } -\int_{\lambda _{i}}^{\lambda }e^{i\theta _{m}\left( \lambda ^{\prime
}\right) -i\theta _{n}\left( \lambda ^{\prime }\right) }d\left[ \frac{%
\left\langle \varphi _{m}(\lambda ^{\prime })\right\vert i\partial _{\lambda
^{\prime }}\left\vert \varphi _{n}(\lambda ^{\prime })\right\rangle }{%
E_{m}(\lambda ^{\prime })-E_{n}(\lambda ^{\prime })-\dot{\lambda}^{\prime }%
\left[ \varepsilon _{m}\left( \lambda ^{\prime }\right) -\varepsilon
_{n}\left( \lambda ^{\prime }\right) \right] }\right]. \nonumber\\
\end{eqnarray}
\end{widetext}
The second term can be dropped out from the first-order correction because
it is a higher order term \cite{APT}. Then, we obtain
\begin{equation}
a_{m}^{\left( 1\right) }\left( t\right) = \sum_{n\neq m}\frac{\left\langle
\varphi _{m}(\lambda )\right\vert i\partial _{\lambda }\left\vert \varphi
_{n}(\lambda )\right\rangle }{E_{m}(\lambda )-E_{n}(\lambda )}e^{i\theta
_{m}\left( \lambda \right) -i\theta _{n}\left( \lambda \right)
}a_{n}^{\left( 0\right) }\left( t_{i}\right).
\end{equation}%
Hence, we obtain the solution up to the first order of ramping velocity as
\begin{equation}
a_{m}\left( t\right) = a_{m}^{\left( 0\right) }\left( t_{i}\right) +\frac{%
\partial \lambda }{\partial t}\sum_{n\neq m}W_{mn}(\lambda )e^{i\theta
_{m}\left( \lambda \right) -i\theta _{n}\left( \lambda \right)
}a_{n}^{\left( 0\right) }\left( t_{i}\right) ,
\end{equation}%
where we have defined
\begin{equation}
W_{mn}(\lambda )=\frac{\left\langle \varphi _{m}(\lambda )\right\vert
i\partial _{\lambda }\left\vert \varphi _{n}(\lambda )\right\rangle }{%
E_{m}(\lambda )-E_{n}(\lambda )}.
\end{equation}

Now we consider that the initial state is the instantaneous ground state of
the initial Hamiltonian, $\left\vert \psi (t_{i})\right\rangle =\left\vert
\varphi _{0}(\lambda _{i})\right\rangle $. That is to say, the initial
condition gives
\begin{equation}
a_{m}^{\left( 0\right) }\left( t_{i}\right) +\left. \frac{\partial \lambda }{%
\partial t}\right\vert _{\lambda _{i}}\sum_{n\neq m}W_{mn}(\lambda
_{i})a_{n}^{\left( 0\right) }\left( t_{i}\right) =\delta _{m,0}.  \label{ICD}
\end{equation}
It is easy to see that $a_{0}^{\left( 0\right) }\left( t_{i}\right) =1$ and $%
a_{m}^{\left( 0\right) }\left( t_{i}\right) =-\frac{\partial \lambda }{%
\partial t}W_{m0}\left( t_{i}\right) $ for $m\neq 0$ satisfy the initial
condition Eq. \ref{ICD} up to the first order. Then at the final time $t_{f}$%
, we can obtain $a_{0}\left( t_{f}\right) \simeq 1$ and for $m\neq 0$,
\begin{align}
&a_{m}\left( t_{f}\right) \simeq \nonumber\\
& \left. \frac{\partial \lambda }{\partial t}%
\right\vert _{\lambda _{f}}W_{m0}(\lambda _{f})e^{i\theta _{m}\left( \lambda
_{f}\right) -i\theta _{0}\left( \lambda _{f}\right) }-\left. \frac{\partial
\lambda }{\partial t}\right\vert _{\lambda _{i}}W_{m0}(\lambda _{i}).
\end{align}
So the wave function at the final time is given by
\begin{align}  \label{perturbation}
&\left\vert \psi (t_{f})\right\rangle =e^{-i\theta _{0}(\lambda
_{f})}\left\vert \varphi _{0}(\lambda _{f})\right\rangle \nonumber \\
&+\left. \frac{\partial \lambda }{\partial t}\right\vert _{\lambda _{f}}\sum\limits_{m\neq
0}e^{-i\theta _{0}\left( \lambda _{f}\right) }W_{m0}(\lambda _{f})\left\vert
\varphi _{m}(\lambda _{f})\right\rangle \nonumber\\
&-\left. \frac{\partial \lambda }{%
\partial t}\right\vert _{\lambda _{i}}\sum\limits_{m\neq 0}e^{-i\theta
_{m}(\lambda _{f})}W_{m0}(\lambda _{i})\left\vert \varphi _{m}(\lambda
_{f})\right\rangle +\cdots.  \notag \\
\end{align}

Using the relation
\begin{equation}
\left\langle \varphi _{m}(\lambda )\right\vert \partial _{\lambda
}\left\vert \varphi _{n}(\lambda )\right\rangle =-\frac{\left\langle \varphi
_{m}(\lambda )\right\vert \partial \hat{H}(\lambda )/\partial \lambda
\left\vert \varphi _{n}(\lambda )\right\rangle }{E_{m}(\lambda
)-E_{n}(\lambda )},
\end{equation}
$W_{mn}(\lambda )$ can be simplified into
\begin{equation}
W_{mn}(\lambda )=-i\frac{\left\langle \varphi _{m}(\lambda )\right\vert \hat{%
V}\left\vert \varphi _{n}(\lambda )\right\rangle }{\left[ E_{m}(\lambda
)-E_{n}(\lambda )\right] ^{2}},
\end{equation}%
where $\hat{V}=\partial \hat{H}\left( \lambda \right) /\partial \lambda $.
Considering the ramping trajectory with $\left. \partial \lambda /\partial
t\right\vert _{\lambda _{i}}=0$, and $\left. \partial \lambda /\partial
t\right\vert _{\lambda _{f}}=\nu $, one obtains
\begin{align}
&\left\vert \psi (t_{f})\right\rangle =e^{-i\theta _{0}(\lambda _{f})}\times\nonumber\\
&\left(
\left\vert \varphi _{0}(\lambda _{f})\right\rangle +\nu \sum\limits_{m\neq
0}W_{m0}(\lambda _{f})\left\vert \varphi _{m}(\lambda _{f})\right\rangle
\right) +\cdots.
\end{align}
Then, measuring an observable $\hat{O}$ at the final time gives
\begin{equation}
\left\langle \hat{O}(t_{f})\right\rangle =\left\langle \varphi _{0}(\lambda
_{f})\right\vert \hat{O}\left\vert \varphi _{0}(\lambda _{f})\right\rangle
+\alpha \nu +\mathcal{O}(\nu ^{2}),  \label{observable}
\end{equation}%
where the first order coefficient $\alpha$ in the expansion Eq. \ref
{observable} is given by
\begin{align}
\alpha =\sum_{m\neq 0}\left\{ \frac{\left\langle \varphi _{0}(\lambda
_{f})\right\vert \hat{O}\left\vert \varphi _{m}(\lambda _{f})\right\rangle
\left\langle \varphi _{m}(\lambda _{f})\right\vert \hat{V}\left\vert \varphi
_{0}(\lambda _{f})\right\rangle }{\left[ E_{m}(\lambda _{f})-E_{0}(\lambda
_{f})\right] ^{2}} \right.\nonumber\\
\left.-\frac{\left\langle \varphi _{0}(\lambda _{f})\right\vert
\hat{V}\left\vert \varphi _{m}(\lambda _{f})\right\rangle \left\langle
\varphi _{m}(\lambda _{f})\right\vert \hat{O}\left\vert \varphi _{0}(\lambda
_{f})\right\rangle }{\left[ E_{0}(\lambda _{f})-E_{m}(\lambda _{f})\right]
^{2}}\right\}.  \label{alpha1}
\end{align}
Note that the instantaneous retarded Green's function at $\lambda_f$ is
given by
\begin{equation}
\mathcal{G}^{R}\left( t,\lambda _{f}\right) =-i\Theta \left( t\right)
\left\langle \varphi _{0}(\lambda _{f})\right\vert \left[ \hat{O}(t),\hat{V}%
(0)\right] \left\vert \varphi _{0}(\lambda _{f})\right\rangle,
\end{equation}
and its spectral presentation in the frequency domain can be written as
\begin{align}
&\mathcal{G}^{R}\left( \omega ,\lambda _{f}\right) =\nonumber\\
&\sum_{m}\left\{ \frac{%
\left\langle \varphi _{0}(\lambda _{f})\right\vert \hat{O}\left\vert \varphi
_{m}(\lambda _{f})\right\rangle \left\langle \varphi _{m}(\lambda
_{f})\right\vert \hat{V}\left\vert \varphi _{0}(\lambda _{f})\right\rangle }{%
\omega -\left[ E_{m}(\lambda _{f})-E_{0}(\lambda _{f})\right] +i0^{+}} \right.\nonumber\\
&\left.-\frac{\left\langle \varphi _{0}(\lambda _{f})\right\vert \hat{V}\left\vert
\varphi _{m}(\lambda _{f})\right\rangle \left\langle \varphi _{m}(\lambda
_{f})\right\vert \hat{O}\left\vert \varphi _{0}(\lambda _{f})\right\rangle }{%
\omega +\left[ E_{m}(\lambda _{f})-E_{0}(\lambda _{f})\right] +i0^{+}}%
\right\} .  \label{GreenFun1}
\end{align}%
Comparing Eq.(\ref{GreenFun1}) and Eq.(\ref{observable}), we arrive at the
result
\begin{equation}
\alpha =i\left. \frac{\partial \mathcal{G}^{R}\left( \omega ,\lambda
_{f}\right) }{\partial \omega }\right\vert _{\omega =0}.
\end{equation}

\section{Examples for the Non-Adiabatic Linear Response \label{example}}

\begin{figure}[t]
\includegraphics[width=0.48\textwidth]{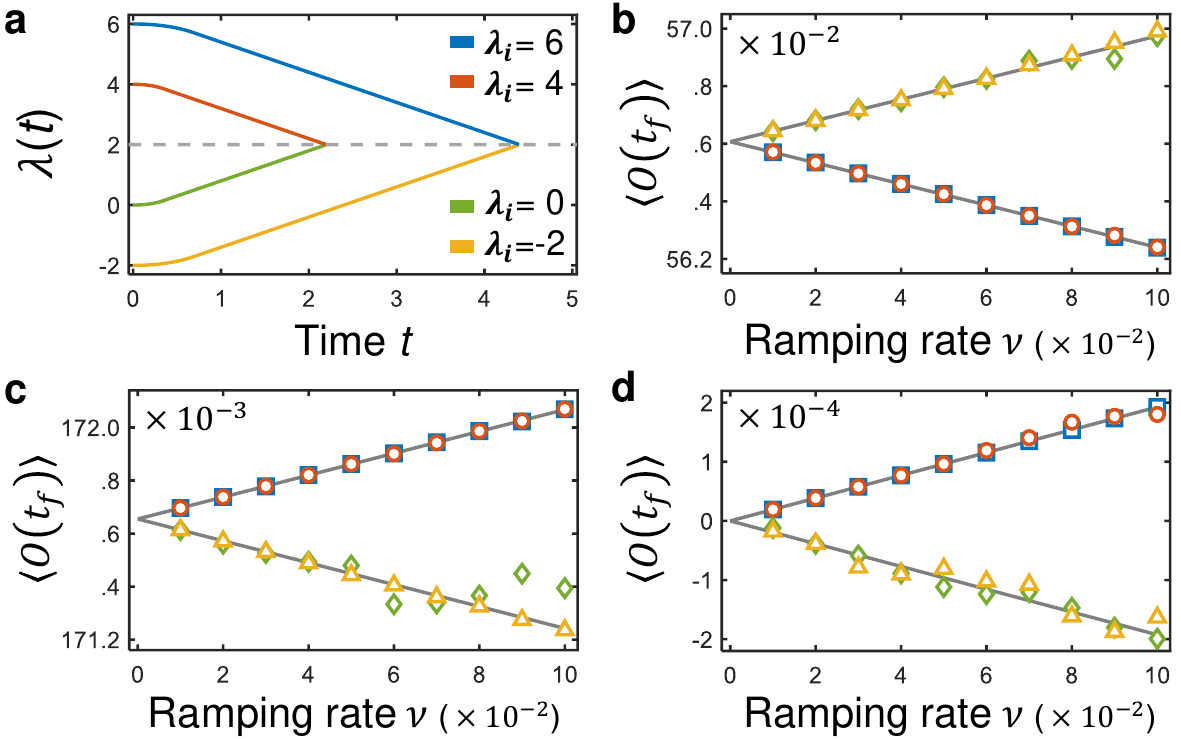}
\caption{(a) Ramping protocol of the parameter $\protect\lambda \left(
t\right)$ given by Eq. \protect\ref{trajactory}. The final value is fixed as
$\protect\lambda _{f}=2$, and initial values are respectively taken as $%
\protect\lambda _{i}=6,4,0,-2$. (b)(c)(d) The measured observable at the
finial time $\left\langle \hat{O}(t_{f})\right\rangle $ as a function of
ramping rate $\protect\nu $ for the transverse Ising model (b), the p-wave
superconductor model (c) and the Bogoliubov model (d). The squares, circles,
diamonds and triangles are results from numerical simulation of the ramping
dynamics with four different initial values, and the solid line is the
corresponding Green's function given by the non-adiabatic linear response
theory. }
\label{examples}
\end{figure}

Now we consider three models as examples to numerically verify the non-adiabatic
linear response theory. The ramping protocol of the parameter $\lambda
\left( t\right) $ is given by (see Fig.~\ref{examples}(a))%
\begin{equation}
\lambda \left( t\right) =\left\{
\begin{array}{c}
\lambda _{i}+a\nu t^{2},\mathrm{\ \ \ \ \ \ \ \ \ \ \ \ \ \ \ \ \ \ }0\leq
t\leq \frac{1}{2a} \\
\lambda _{i}+\frac{\nu }{4a}+\nu \left( t-\frac{1}{2a}\right) ,\mathrm{\ \ \
\ \ \ \ \ \ }t>\frac{1}{2a}%
\end{array}%
,\right.  \label{trajactory}
\end{equation}%
where $a=\frac{5\nu }{2\left( \lambda _{f}-\lambda _{i}\right) }$. This
protocol gives a smooth $\lambda(t)$ curve that satisfies $\left. \partial
\lambda /\partial t\right\vert _{\lambda _{i}}=0$ and $\left. \partial
\lambda /\partial t\right\vert _{\lambda _{f}}=\nu $. We fix the final
parameter as $\lambda _{f}=2$, and start with four different initial
parameters as $\lambda _{i}=6,4,0,-2$. For each initial $\lambda $, we use
different ramping rate $\nu $. We numerically simulate the ramping dynamics
and then compare the results with the prediction of the non-adiabatic linear
response theory.

Here we consider three different models. The first mode is the quantum Ising
model with external magnetic fields, whose Hamiltonian is given by
\begin{align}
\hat{H}_{1}(\lambda )=&-J\sum_{i}\sigma _{i+1}^{z}\sigma _{i}^{z}-\lambda
(t)h_{x}\sum_{i}\sigma _{i}^{x}\nonumber\\
&-h_{y}\sum_{i}\sigma_{i}^{y}-h_{z}\sum_{i}\sigma _{i}^{z}.
\end{align}%
The ramping term is an external field along $\hat{x}$ with $\hat{V}%
=h_{x}\sum_{i}\sigma _{i}^{x}$, and the measurement operator $\hat{O}$ is
taken as spin along $\hat{y}$ with $\hat{O}=\sigma _{i}^{y}$. The numerical results are plotted in Fig.~\ref
{examples} (b) with system length $L=8$. Here $J=1$ is set as the energy unit and $1/J$ is taken as
the time unit ($\hbar =1$). In the plot we set $h_{x}=1$,$h_{y}=2$ and $h_{z}=1$. The second mode is a $p$-wave superconductor induced by the
proximity effect, whose Hamiltonian is given by
\begin{align}
\hat{H}_{2}(\lambda )=&\lambda (t)\sum_{\mathbf{k},\sigma }\epsilon _{\mathbf{%
k}}\hat{c}_{\mathbf{k},\sigma }^{\dagger }\hat{c}_{\mathbf{k},\sigma
}\nonumber\\&+\Delta _{0}\sum_{\mathbf{k}}\left[ \left( k_{x}-ik_{y}\right) \hat{c}_{%
\mathbf{k},\uparrow }^{\dagger }\hat{c}_{-\mathbf{k},\downarrow }^{\dag
}+h.c.\right] ,
\end{align}%
where $\epsilon _{\mathbf{k}}=-2t_{h}(\cos (k_{x})+\cos (k_{y}))$. The
ramping term is the kinetic energy term with $\hat{V}=\sum_{\mathbf{k}%
,\sigma }\epsilon _{\mathbf{k}}\hat{c}_{\mathbf{k},\sigma }^{\dagger }\hat{c}%
_{\mathbf{k},\sigma }$, and the measurement operator $\hat{O}$ is taken as
the paring order $\hat{O}_{2}=\frac{1}{2}\left( \hat{c}_{\mathbf{k},\uparrow
}^{\dagger }\hat{c}_{-\mathbf{k},\downarrow }^{\dag }+\hat{c}_{-\mathbf{k}%
,\downarrow }\hat{c}_{\mathbf{k},\uparrow }\right) $. Since different
momentum $\mathbf{k}$ are decoupled in this model, we focus on the specific
momentum with $\mathbf{k}=(\pi ,\pi )$. The numerical results are plotted in
Fig.~\ref{examples}(c). Here $t_{h}=1$ is set as the energy unit and $%
1/t_{h}=1$ is taken as the time unit. In the plot we set $\Delta _{0}=1
$. The third model is the Bogoliubov model of the Bose-Einstein condensates,
whose Hamiltonian is given by
\begin{align}
\hat{H}_{3}(\lambda )=&\sum_{\mathbf{k}}\left( \left( \lambda +\lambda
_{0}\right) \epsilon _{\mathbf{k}}+gn\right) a_{\mathbf{k}}^{\dagger }a_{%
\mathbf{k}}\nonumber\\
&+gn\sum_{\mathbf{k}}\left( \hat{a}_{\mathbf{k}}^{\dagger }\hat{a}%
_{-\mathbf{k}}^{\dag }+h.c.\right) ,
\end{align}%
where $\epsilon _{\mathbf{k}}=\mathbf{k}^{2}/2m$. We have taken $\lambda
_{0}=5$ to ensure $\lambda +\lambda _{0}$ is always positive, such that the
excitation is dynamical stable throughout the entire ramping proces. Unlike
the above two models, this model is always gapless. The ramping term is also
the kinetic energy term with $\hat{V}=\sum_{\mathbf{k}}\epsilon _{\mathbf{k}%
}a_{\mathbf{k}}^{\dagger }a_{\mathbf{k}}$, and the we measure the response
of $\hat{O}_{3}=-i\left( \hat{a}_{\mathbf{k}_{r}}^{\dagger }\hat{a}_{-%
\mathbf{k}_{r}}^{\dag }-\hat{a}_{-\mathbf{k}_{r}}\hat{a}_{\mathbf{k}%
_{r}}\right) $, where $\mathbf{k}_{r}$ is a given momentum. The results are
plotted in Fig.~\ref{examples}(d). Here $E_{r}=\mathbf{k}_{r}^{2}/2m$ is set
as the energy unit and $1/E_{r}$ is taken as the time unit ($\hbar =1$). In
the plot we set $gn=5$. In all these three examples, we can see from
Fig. \ref{examples}(b-d) that the linear slope is independent of the ramping
trajectories, and the slope is consistent with the Green's function given by
the solid lines.

\section{Applicable Condition of the Non-Adiabatic Linear Response \label{condition}}

This theory concerns the first-order expansion in term of the ramping
velocity. Therefore, the validity conditions of our theory are two folds.
First, the first order coefficient does not vanish. Secondly, the high order
coefficients do not diverge. As long as these two conditions are satisfied,
there is always a regime where the linear expansion is valid, although the
linear regime depends on the ratio between the high order and the first
order coefficients.

First, we discuss when the first order coefficient $\alpha$ vanishes. It is
obvious from Eq. 3 of the main text that $\alpha$ vanishes if $\hat{O}=\hat{V}$. If
$\hat{O}\neq \hat{V}$, $\alpha$ also vanishes if there exists an
anti-unitary operator $\hat{\Theta}=\hat{U}\hat{K}$, where $\hat{U}$ is a
unitary operator and $\hat{K}$ is taking complex conjugate, such that
operators $\hat{O},\hat{V}$ and instantaneous eigen-states are all invariant
under this anti-unitary transformation, i.e.
\begin{eqnarray}
\hat{\Theta}\hat{V}\hat{\Theta}^{-1} &=&\hat{V}, \\
\hat{\Theta}\hat{O}\hat{\Theta}^{-1} &=&\hat{O}, \\
\hat{\Theta}\left\vert \varphi _{n}(\lambda _{f})\right\rangle &=&\left\vert
\varphi _{n}(\lambda _{f})\right\rangle .
\end{eqnarray}%
The proof is following. For any given anti-unitary operator $\hat{\Theta}=%
\hat{U}\hat{K}$, we have \cite{Sakurai}
\begin{equation}
\left\langle \varphi _{m}(\lambda _{f})\right\vert \hat{V}\left\vert \varphi
_{n}(\lambda _{f})\right\rangle =\left\langle \tilde{\varphi}_{n}(\lambda
_{f})\right\vert \hat{\Theta}\hat{V}\hat{\Theta}^{-1}\left\vert \tilde{%
\varphi}_{m}(\lambda _{f})\right\rangle .
\end{equation}%
where $\left\vert \tilde{\varphi}_{n}(\lambda _{f})\right\rangle =\hat{\Theta%
}\left\vert \varphi _{n}(\lambda _{f})\right\rangle $. If $\hat{\Theta}\hat{V%
}\hat{\Theta}^{-1}=\hat{V}$ and $\left\vert \tilde{\varphi}_{n}(\lambda
_{f})\right\rangle =\left\vert \varphi _{n}(\lambda _{f})\right\rangle $,
one obtains
\begin{equation}
\left\langle \varphi _{m}(\lambda _{f})\right\vert \hat{V}\left\vert \varphi
_{n}(\lambda _{f})\right\rangle =\left\langle \varphi _{n}(\lambda
_{f})\right\vert \hat{V}\left\vert \varphi _{m}(\lambda _{f})\right\rangle.
\label{Vexg}
\end{equation}
And the same holds for the operator $\hat{O}$. Substituting this identity
into Eq. 3 of the main text, one finds that $\alpha =0$. Thus, our theory is valid
when such an anti-unitary symmetry does not exist.

Secondly, we look into the higher order terms. Following the expansion
discussed above, we can obtain
\begin{align}
&a_{n}\left( t_{f}\right) =\nu \frac{\left\langle \varphi _{n}(\lambda
_{f})\right\vert i\partial _{\lambda }\left\vert \varphi _{0}(\lambda
_{f})\right\rangle }{E_{n}(\lambda _{f})-E_{0}(\lambda _{f})}  \nonumber\\
&+\nu
^{2}\sum_{m\neq 0,n}\frac{\left\langle \varphi _{n}(\lambda _{f})\right\vert
i\partial _{\lambda }\left\vert \varphi _{m}(\lambda _{f})\right\rangle }{%
E_{m}(\lambda _{f})-E_{0}(\lambda _{f})}\frac{\left\langle \varphi
_{m}(\lambda _{f})\right\vert i\partial _{\lambda }\left\vert \varphi
_{0}(\lambda _{f})\right\rangle }{E_{n}(\lambda _{f})-E_{0}(\lambda _{f})}\nonumber\\
&+\dots
\end{align}%
Here we focus on the second term as an example, and we can replace the
summation as an integration over the energy, which leads to
\begin{align}
&\sum_{m\neq 0,n}\frac{\left\langle \varphi _{n}(\lambda _{f})\right\vert
i\partial _{\lambda }\left\vert \varphi _{m}(\lambda _{f})\right\rangle }{%
E_{m}(\lambda _{f})-E_{0}(\lambda _{f})}\frac{\left\langle \varphi
_{m}(\lambda _{f})\right\vert i\partial _{\lambda }\left\vert \varphi
_{0}(\lambda _{f})\right\rangle }{E_{n}(\lambda _{f})-E_{0}(\lambda _{f})}
\nonumber\\
&\lesssim \frac{w^2 }{E_{n}(\lambda _{f})-E_{0}(\lambda _{f})}
\int_{0}^{\Lambda }d\varepsilon \frac{\rho \left( \varepsilon \right) }{%
\varepsilon },  \label{second.order}
\end{align}%
where we have assumed the dimensionless matrix element $\left\langle \varphi
_{n}(\lambda )\right\vert i\partial _{\lambda }\left\vert \varphi
_{m}(\lambda )\right\rangle$ is bounded by $w$. Here $\Lambda $ is a high
energy cutoff, and $\rho \left( \varepsilon \right) $ is the
density-of-state. For a gapped system, the integral in Eq. (\ref%
{second.order}) is finite. For a gapless system, we assume that the low
energy density of states behaves like $\rho \left( \varepsilon \right) \sim
\varepsilon ^{\gamma }$, and when $\gamma > 0$, the integral is also
finite. That is to say, as long as the low-energy density of states
vanishes at $\varepsilon\rightarrow 0$, the second order contribution is finite. Similar arguments can
be applied to higher order terms. When these higher order terms are
finite, the convergent radius of this perturbation series is finite and the
perturbation expansion is valid.

Following our derivations, if we now consider the population on the excited states as these references did, we obtain, to the linear order of $\delta$, 
\begin{align}
&n_\text{ex}=\sum_{n\neq 0}|a_{n}(t)|^{2} =\sum_{n\neq 0}\nu ^{2}\left\vert \frac{
\left\langle \varphi _{n}(\lambda _{f})\right\vert i\partial _{\lambda
}\left\vert \varphi _{0}(\lambda _{f})\right\rangle }{E_{n}(\lambda
_{f})-E_{0}(\lambda _{f})}\right\vert ^{2} \nonumber \\
& \lesssim \nu ^{2}w^{2}\int_{0}^{\Lambda }d\varepsilon \left\vert \frac{
\rho \left( \varepsilon \right) }{\varepsilon }\right\vert ^{2}.
\label{excitations}
\end{align}
If $\gamma\leqslant 0$, the second-order coefficient in the expansion diverges, and the integral in Eq. (\ref{excitations}) should also diverge. The divergent linear coefficient in $n_\text{ex}$ versus $\delta$ implies a non-analytical dependence on $\delta$, consistent with the conclusion in the previous literatures \cite{Polkovnikov2005,Polkovnikov2008}.

\section{Time Sequence of Parametrical Ramping}\label{Sec:time}
In our experiments, we need to eliminate the influences of the non-zero time derivative of trap depth at the start point of ramping.
Therefore, the time sequence of ramp is smoothed such that the initial time derivative vanishes, that is, $\left.{\partial V\over \partial t}\right|_{V=V_i} = 0$. Here, $V$ is the trap depth of the optical lattices. We use the combinations of exponential functions and linear functions to realize such a smoothing ramping trajectory. Initially, the slope of the ramp grows gradually and once it reaches the target value of the time derivative ${\partial V\over \partial t} = \nu$, the ramping function becomes linear until reaching the final trap depth $V_f$. As a piecewise function, the ramping trap depth can be written as
\begin{equation}
V(t)=\left\{
\begin{aligned}
V_i + A(e^{t/\tau}-t/\tau - 1), \ t \leq \tau \\
V_i + \nu(t-\tau) + A(e-2), \ t > \tau
\end{aligned}
\right.
\label{tse}
\end{equation}
where the time constant $\tau$ is set to be larger than the tunneling time scale $\hbar/J$ at the initial states and $A$ depicts the duration of the smoothing sequence. In order to guarantee the function and its first-order derivative to be smooth, it requires $\tau$, $A$, and $\nu$ to satisfy $\nu\tau = A(e-1)$. In Table.~\ref{parameter}, we list the trap depth ramping parameters used in our experiments.

\begin{table*}[htb]
\begin{tabular}{cccccccc}
\hline
\multirow{2}{*}{$V_f(E_r)$} & \multirow{2}{*}{$V_i(E_r)$} & \multirow{2}{*}{$A(E_r)$} & \multicolumn{2}{c}{$\tau$} & \multirow{2}{*}{$\hbar/J$(ms)} & \multicolumn{2}{c}{$J\tau/\hbar$} \\
                        &                         &                        & min(ms)    & max(ms)   &                                             & min           & max          \\
                        \hline
\multirow{2}{*}{11}     & 15                      & -2                      & 11.46      & 34.37     & 11.79                                       & 0.972         & 2.915        \\
                        & 17                      & -2                      & 11.46      & 34.37     & 17.49                                       & 0.655         & 1.965        \\
                        \hline
\multirow{2}{*}{13}     & 17                      & -2                      & 11.46      & 34.37     & 17.49                                       & 0.655         & 1.965        \\
                        & 19                      & -2                      & 11.46      & 34.37     & 25.56                                       & 0.448         & 1.345        \\
                        \hline
\multirow{6}{*}{15}     & 5                       & 4                      & 22.91      & 68.73     & 1.17                                        & 19.582        & 58.745       \\
                        & 11                      & 4                      & 22.91      & 68.73     & 5.05                                        & 4.537         & 13.610       \\
                        & 17                      & -2                      & 11.46      & 34.37     & 17.49                                       & 0.655         & 1.965        \\
                        & 18                      & -2                      & 11.46      & 34.37     & 21.19                                       & 0.541         & 1.622        \\
                        & 19                      & -2                      & 11.46      & 34.37     & 25.56                                       & 0.448         & 1.345        \\
                        & 20                      & -5                      & 28.64      & 85.91     & 30.73                                       & 0.932         & 2.796        \\
                        \hline
\multirow{2}{*}{17}     & 11                      & 2                      & 11.46      & 34.37     & 5.05                                        & 2.268         & 6.805        \\
                        & 13                      & 2                      & 11.46      & 34.37     & 7.80                                        & 1.469         & 4.406        \\
                        \hline
\multirow{2}{*}{19}     & 13                      & 2                      & 11.46      & 34.37     & 7.80                                        & 1.469         & 4.406        \\
                        & 15                      & 2                      & 11.46      & 34.37     & 11.79                                       & 0.972         & 2.915        \\
                        \hline
\multirow{2}{*}{21}     & 15                      & 2                      & 11.46      & 34.37     & 11.79                                       & 0.972         & 2.915        \\
                        & 17                      & 2                      & 11.46      & 34.37     & 17.49                                       & 0.655         & 1.965
\end{tabular}
\caption{\label{parameter}\textbf{Trap depth ramping.}
We list the corresponding $A$ and $\hbar/J$ which are fixed values for each combination of $V_f$ and $V_i$.
For different ramping velocities $\nu$, we apply different values of $\tau$ and list the maximum and minimum ones. The smaller $\tau$ corresponds to a faster ramp with speed $|\nu|=0.3E_r$/ms, and the larger $\tau$ corresponds to a slower speed with $|\nu|=0.1E_r$/ms. }
\end{table*}

\section{Day-to-day Drift and Lattice Heating}\label{Sec:drift}
\begin{figure}
\includegraphics[width=\linewidth]{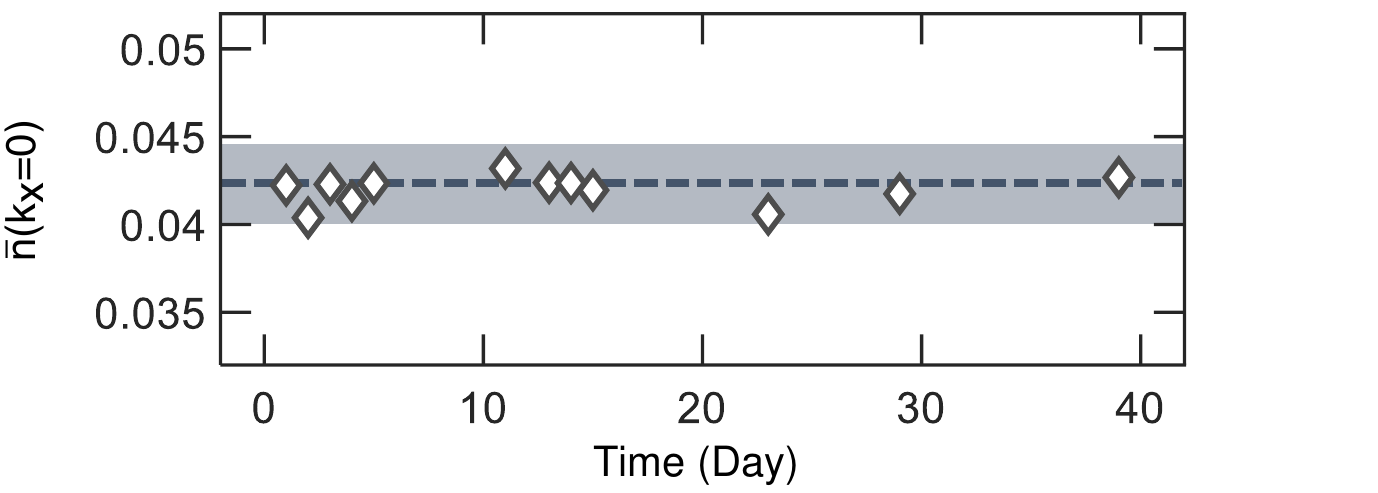}
  \vspace{-0.25cm}
  \caption{ { \textbf{Day-to-day drift in measurement of $\bar{n}(k_x=0)$.} We measure $\bar{n}(k_x=0)$ of the steady states by holding the system at $15E_r$ for $35$ms at each day. By summarizing the measurements of different days, we obtain a distribution of $\bar{n}(k_x=0)$ at a mean value 0.0422 and a standard deviation 0.0021 averaged which is averaged by $300$ measurements in 40 days. It shows a day-to-day drift around $10\%$ for $\bar{n}(k_x=0)$ between different days.
    \label{Drift} \vspace{-0.25cm}
 }}
\end{figure}

In the limit of $\nu \to 0$, we should obtain the same $\bar{n}(k_x=0)$ for a given $V_f$ with different $V_i$, which recover the adiabatic limit. However, there is a small discrepancy between different data sets in our experiments. This is due to the day-to-day drift in our system. To confirm this, here we measure the same observable $\bar{n}(k_x=0)$ of steady states at $V=15E_r$ in different days, and the results are shown in Fig.~\ref{Drift}. We find that, within one standard deviation confidence, the fluctuation covers the discrepancy in our measurements.
We think that this drift mainly arises from slight differences of system vacuum pressure, temperatures and humidities on different days. This day-to-day drift only changes the intercepts of the linear results and does not hurt the slopes, because data for each curve with a given pair of initial $V_i$ and final $V_f$ is taken within one day to avoid the systematic drifts.

\begin{figure}[htb]
\includegraphics[width=\linewidth]{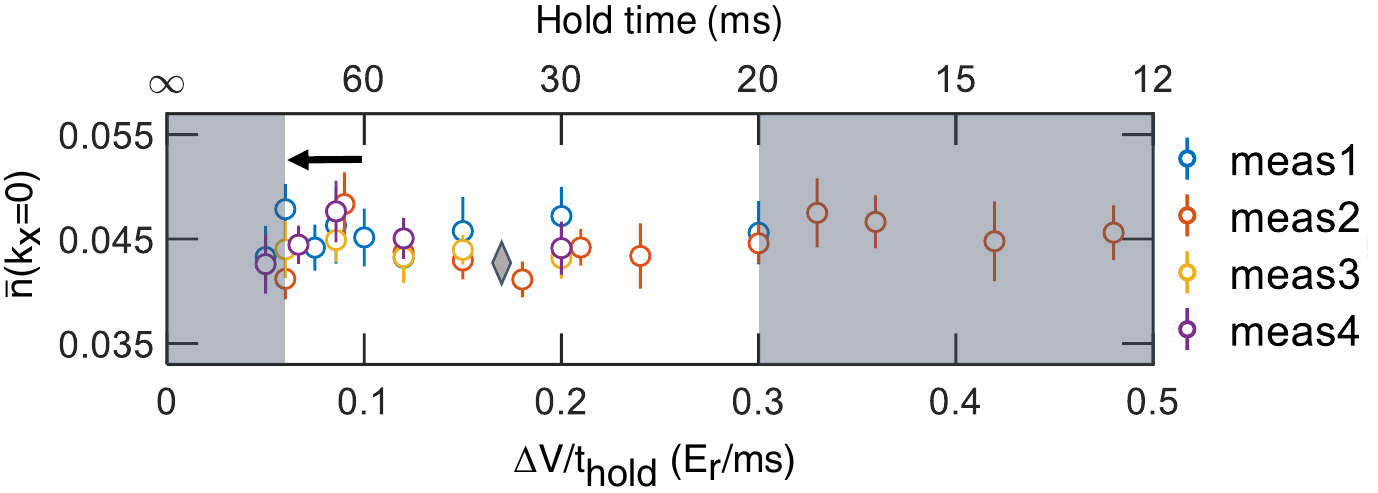}
  \vspace{-0.25cm}
  \caption{ \textbf{$\bar{n}(k_x=0)$ versus the holding time $t_{hold}$.} Here we list four sets of different holding measurements at $V=15E_r$. The horizontal axis is labeled by the holding time. To give an intuitive comparison with the ramping velocity $\nu$, we use the typical change of the trap depth $\Delta V=6E_r$, to plot a second label of the horizontal axis $\Delta V/t_{hold}$, which can be compared with the ramping velocity $\nu$. The unshadowed area corresponds to the region of $\nu$ used in our experiments.
 The data proves that heating does not show significant effects here. The grey diamond denotes the data $\bar{n}(k_x=0)=0.042(2)$ obtained from Fig.~\ref{Drift} with a holding time $t_{hold}=35$ ms, which is consistent with the measurements here.\label{heating} \vspace{-0.25cm}}
\end{figure}

Besides calibrating the day-to-day drifts, we also calibrate the heating from the optical lattices. Here we vary the holding time $t_{hold}$ from $10$~ms to $120$~ms after adiabatically ramping to the steady states at $15E_r$, in order to check whether the linear dependence will be hurt by the heating.
In Fig.~\ref{heating}, the measured $\bar{n}(k_x=0)$ doesn't show an explicit dependence on the holding time $t_{hold}$.
Therefore, we verify that the heating effect is negligible during the time scale of our experiments and does not affect our experimental results.

\section{Fitting the Quasi-momentum Profiles in the First Brillouin Zone}\label{Sec:fitting}
We divide the quasi-momentum profiles into three parts. A central Lorentzian peak corresponds to the coherent part, a Gaussian wing corresponds to the thermal atoms, and a flat plateau corresponds to the incoherent parts due to Mott insulators. Therefore, the entire fitting function is written as
\begin{equation}
n(q_x) = \frac{A}{(q_x-q_0)^2 + (\Gamma/2)^2} + B\cdot \exp(-\frac{(q_x-q_0)^2}{2w_0^2}) + C.
\label{all}
\end{equation}
Here $q_x$ is the quasi-momentum, $q_0$ characterizes the zero-momentum point in raw data which is obtained via fitting, and $\Gamma$ and $\omega_0$ characterize the width of the Lorentzian and Gaussian shapes. Thus, the peak value of the three-components distribution is $n_{q_x=q_0} = 4A/\Gamma^2 + B + C$.

\begin{figure}[htb]
\includegraphics[width=0.8\linewidth]{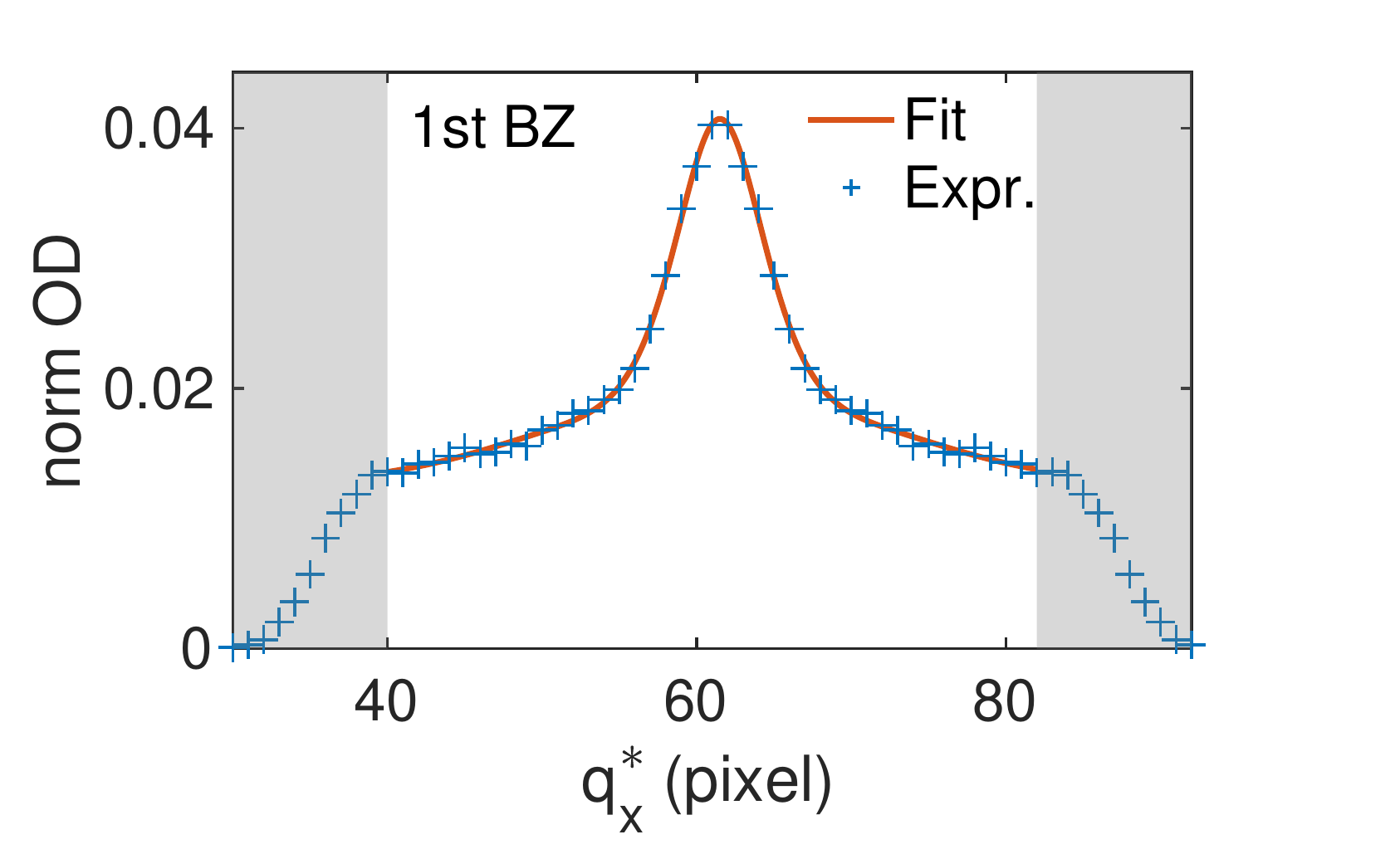}
  \vspace{-0.25cm}
  \caption{ { \textbf{The three-component fitting:}
 The blue crosses denote experimental data.
 The unshadowed region labels the first Brillouin zone.
 The horizontal axis is labeled by the pixel of our imaging camera, and the vertical axis is labeled by the normalized optical density (OD). Here the optical density is normalized by the total atom number to avoid the loading fluctuations in each measurement.
 The red solid line is the three-component fitting curve.
  \label{fitting} \vspace{-0.25cm}
 }}
\end{figure}

For each raw data, we symmetrize the profile by adding its mirrored version around the geometric center to avoid asymmetric systematic errors. In Fig.~\ref{fitting}, we show one example of the symmetrized data and the fitting function. The three-component fitting model fits nicely with our measured data. With such fitting, we are able to extract out quasi-momentum distribution $\bar{n}(k_x)$ in the first Brillouin zone. This enables us to eliminate statistical fluctuations of each single data point, and leads to a more robust analysis of $\bar{n}(k_x)$ versus $\nu$.

\begin{figure*}[htb]
\includegraphics[width=0.9\linewidth]{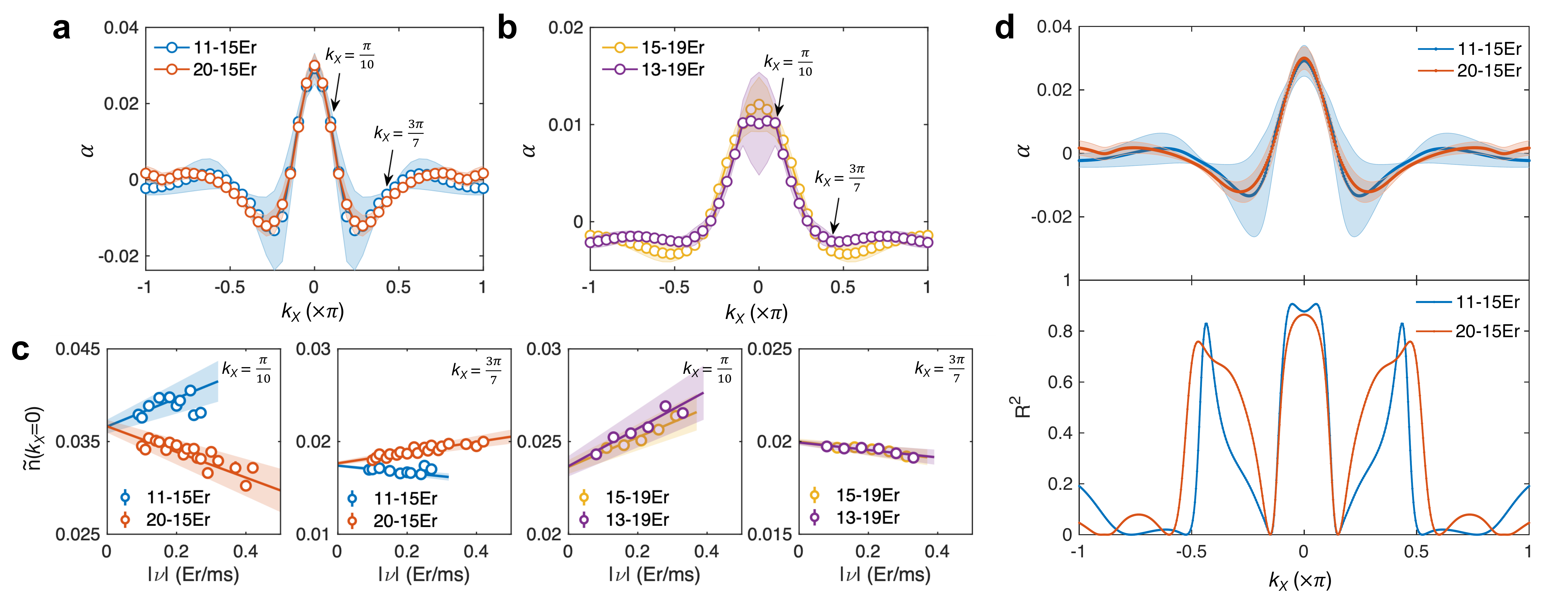}
  \vspace{-0.25cm}
  \caption{ { \textbf{The linear dependence in the first Brillouin zone.} \textbf{a} and \textbf{b}
  are the plots of Fig.~3\textbf{C} and \textbf{D} in the main text.
  \textbf{c} shows $\bar{n}(k_x)$ versus $|\nu|$ for different ramps, initial trap depth and final trap depth at $k_x = \frac{\pi}{10},\ \frac{3\pi}{7}$. \textbf{d} shows $\alpha$ versus $k_x$ for $V_f=15E_r$. The blue solid line corresponds to the ramp from 11 to 15$E_r$, and the red solid line corresponds to the ramp from 20 to 15$E_r$. The shadow areas correspond to the one standard deviation confident region. Here we also show the r-square of fitting $\bar{n}(k_x)$ versus $k_x$. There is a sign-flipping point of $\alpha$. At this point, the slope $\alpha\rightarrow 0$ vanishes and the data points are distributed purely by the experimental and statistical noises. The linearity of fitting becomes unstable, and this will artificially lower the r-square.}
    \label{linearDependence} \vspace{-0.25cm}
 }
\end{figure*}

After fitting the quasi-momentum profiles, we revisit the results presented in Fig.~3\textbf{C} and \textbf{D} in the main text (Fig.~\ref{linearDependence} \textbf{a} and \textbf{b} here).
We choose particular quasi-momentum $k_x = \frac{\pi}{10},\ \frac{3\pi}{7}$ in each graph and plot $\bar{n}(k_x)$ versus the ramping velocity $\nu$ in Fig.~\ref{linearDependence}\textbf{c}.
We see a linear dependence of $\bar{n}(k_x)$ on $\nu$ for non-zero quasi-momenta.
Besides these two momenta at $k_x = \frac{\pi}{10},\ \frac{3\pi}{7}$, we obtain the linear slope $\alpha$ for each quasi-momentum $k_x$ in the first Brillouin zone (Fig.~\ref{linearDependence}~\textbf{d}).
The slopes $\alpha$ with the same final trap depth, obtained via two different ramping trajectories, are consistent with each other within one standard deviation confidence.
In Fig.~\ref{linearDependence} \textbf{d}, we also plot the r-square value for the linear fitting at each $k_x$ to show the fidelity of the linear fit. The sign of $\alpha$ flips at around $k_x = \pi/7$. Away from this sign-flip point, the r-square reaches above $0.75$ which supports the linear dependence.

\section{Simplifying the Correlation Function in the Bose-Hubbard model \label{BHM}}

Now we apply the non-adiabatic linear response theory to the ramping process
of the Bose-Hubbard model in an optical lattice. The Hamiltonian and the
ramping protocol are given by
\begin{equation}
\hat{H}\left( \lambda \right) =-J\sum_{\left\langle ij\right\rangle }\left(
\hat{a}_{i}^{\dag }\hat{a}_{j}+h.c.\right) +\sum_{i} \frac{U \left( \lambda
\right)}{2} \hat{n}_{i} \left( \hat{n}_{i} - 1 \right) ,
\end{equation}%
where $\hat{V}=\partial \hat{H}\left( \lambda \right) /\partial \lambda
=\sum_{i}\hat{n}_{i}\left( \hat{n}_{i}-1\right) =\frac{1}{N_{\mathrm{s}}}%
\sum_{\mathbf{pk}_{2}\mathbf{k}_{1}}\hat{a}_{\mathbf{p+k}_{1}}^{\dag }\hat{a}%
_{\mathbf{p-k}_{1}}^{\dag }\hat{a}_{\mathbf{p-k}_{2}}\hat{a}_{\mathbf{p+k}%
_{2}}$, and $N_{\mathrm{s}}$ is the total number of the optical lattice
site. The observable in the experiment is the momentum distribution $\hat{O}
=\hat{a}_{\mathbf{k}}^{\dag }\hat{a}_{\mathbf{k}}$. Therefore, the
corresponding retarded Green's function is expressed as
\begin{eqnarray}
i\mathcal{G}^{R}\left( t,\lambda _{f}\right)& =& \frac{\Theta \left( t\right)
}{2N_{\mathrm{s}}}\sum_{\mathbf{pk}_{2}\mathbf{k}_{1}}\left\langle \left[
\hat{a}_{\mathbf{k}}^{\dag }\left( t\right) \hat{a}_{\mathbf{k}}\left(
t\right) ,\hat{a}_{\mathbf{p+k}_{1}}^{\dag }(0)\right.\right. \nonumber \\
& &\left.\left.\hat{a}_{\mathbf{p-k}_{1}}^{\dag }(0)\hat{a}_{\mathbf{p-k}_{2}}(0)\hat{a}_{\mathbf{p+k}_{2}}(0)%
\right] \right\rangle .
\end{eqnarray}%
To evaluate this (real time) retarded Green's function, as usual, we first
calculate the imaginary time correlation function $\mathsf{G}(\tau)$,
\begin{eqnarray}
\mathsf{G}(\tau)& =& - {1\over 2N_s}\sum_{\mathbf{pk}_{2}\mathbf{k}_{1}} \left \langle
T_{\tau} \hat{a}_{\mathbf{k}}^{\dag}(\tau) \hat{a}_{\mathbf{k}}(\tau -
0^{+})a_{\mathbf{p+k}_{1}}^{\dag}(0^{+})
 \right.\nonumber  \\
& &\left.
 \hat{a}_{\mathbf{p-k}%
_{1}}^{\dag}(0^{+}) \hat{a}_{\mathbf{p-k}_{2}}(0) \hat{a}_{\mathbf{p+k}%
_{2}}(0) \right \rangle \, ,
\end{eqnarray}
where $T_{\tau}$ is time ordering operator and certain time arguments have
been shifted infinitesimally to make the expression unambiguous, and then
perform an analytic continuation to real time.

\begin{figure*}[htbp]
\centering
\includegraphics[width=0.85\linewidth]{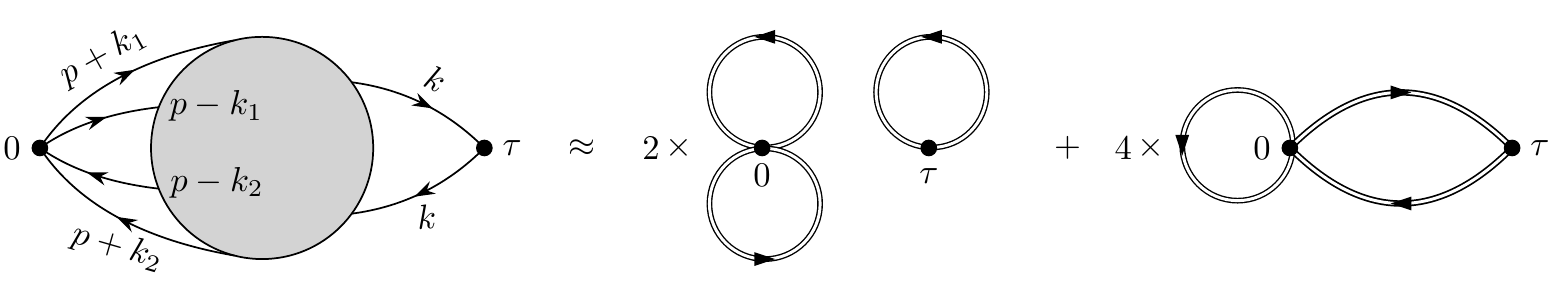}
\caption{Our approximation scheme is to replace the six-point correlator
with pairs of full single-particle correlators (shown as double lines). The
prefactor $2$ and $4$ are the multiplicity of the corresponding diagrams.}
\label{fig:wickcontraction}
\end{figure*}

To evaluate this six-point correlator, we employ the Wick contraction to
approximate this multiple-point correlator into a product of two-point
correlation functions, and this approximation includes the full interaction
effects in the level of two-point correlation and ignores the vertex
correction (see Fig.~\ref{fig:wickcontraction}). With this approximation, one obtains
\begin{equation}
\mathsf{G}(\tau )\approx\mathsf{G}^W(\tau )=-2\bar{n}\langle T_{\tau }\hat{a}_{k}^{\dagger }(\tau
)a_{k}(0)\rangle \langle T_{\tau }\hat{a}_{k}(\tau )a_{k}^{\dagger
}(0)\rangle \,.  \label{eq:G_tau}
\end{equation}%
where $\bar{n}=N/N_{\mathrm{s}}$ is the filling factor. From the K\"{a}ll\'{e%
}n-Lehmann spectral representation, it is easy to show
\begin{align}
\langle T_{\tau }\hat{a}_{k}^{\dagger }(\tau )a_{k}(0)\rangle & =\frac{1}{%
\beta }\sum_{n}e^{-i\omega _{n}\tau }\int \frac{\mathcal{A}(\mathbf{k},\omega )}{%
i\omega _{n}+\omega }d\omega \,, \\
\langle T_{\tau }\hat{a}_{k}(\tau )a_{k}^{\dagger }(0)\rangle & =\frac{1}{%
\beta }\sum_{n}e^{-i\omega _{n}\tau }\int \frac{-\mathcal{A}(\mathbf{k},\omega )}{%
i\omega _{n}-\omega }d\omega \,,
\end{align}%
where $\mathcal{A}(\mathbf{k},\omega )$ is the single-particle spectral function.
Substituting these two relations into Eq.~\eqref{eq:G_tau} and then performing
the Fourier transformation, we end up with
\begin{equation}
\mathsf{G}(i\omega _{n})=2\bar{n}\int d\omega _{1}d\omega _{2}\mathcal{A}(\mathbf{k}%
,\omega _{1})\mathcal{A}(\mathbf{k},\omega _{2})\frac{f_{B}(\omega _{1})-f_{B}(\omega
_{2})}{i\omega _{n}+\omega _{2}-\omega _{1}}
\end{equation}%
where $f_{B}(\omega )=1/(e^{\beta \left( \omega -\mu \right) }-1)$ is the
Bose distribution function. One can now perform the analytic
continuation, $i\omega _{n}\rightarrow \omega +i0^{+}$, to arrive at the
expression of the retarded Green's function $\mathcal{G}^{R}$ in the
frequency domain as
\begin{equation}
\mathcal{G}^{R}(\omega )=2\bar{n}\int d\omega _{1}d\omega _{2}\mathcal{A}(\mathbf{k}%
,\omega _{1})\mathcal{A}(\mathbf{k},\omega _{2})\frac{f_{B}(\omega _{1})-f_{B}(\omega
_{2})}{\omega +\omega _{2}-\omega _{1}+i0^{+}}\,.
\end{equation}%
It is now straightforward to evaluate the slope $\alpha $ as
\begin{eqnarray}
\alpha & =&i\left. \frac{\partial \mathcal{G}^{R}(\omega ,\lambda _{f})}{%
\partial \omega }\right\vert _{\omega =0}
\nonumber\\
&=&2\bar{n}\int d\omega _{1}d\omega
_{2}\mathcal{A}(\mathbf{k},\omega _{1})\mathcal{A}(\mathbf{k},\omega _{2})\nonumber \\
& &\times\frac{f_{B}(\omega
_{2})-f_{B}(\omega _{1})}{\omega _{2}-\omega _{1}}\frac{i}{\omega
_{2}-\omega _{1}+i0^{+}}\nonumber \\
& =&2\pi \bar{n}\int d\omega \mathcal{A}^{2}(\mathbf{k},\omega )f_{B}^{\prime }(\omega
) \nonumber \\
&=&4\pi \bar{n}\int d\omega \mathcal{A}(\mathbf{k},\omega )\frac{d\mathcal{A}(\mathbf{k}
,\omega )}{d\omega}f_{B}(\omega ),
\end{eqnarray}
where the last step follows from the integration by parts.
Deeply in the superfluid or the Mott insulator phase, there exists well-defined
quasi-particles and the spectral function behaves as
\begin{equation}
\mathcal{A}(\mathbf{k},\omega )\sim \frac{\Gamma _{\mathbf{k}}}{(\omega -\epsilon _{%
\mathbf{k}})^{2}+\Gamma _{\mathbf{k}}^{2}}\,,
\end{equation}%
where $\varepsilon _{\mathbf{k}}$ is the quasi-particle dispersion. When the
quasi-particle lifetime is long enough, $\Gamma _{\mathbf{k}}\rightarrow 0$
and $k_{\text{B}}T\gg \Gamma _{\mathbf{k}}$. Then, $f_{\mathrm{B}}(\omega )$
can be taken as a constant in the energy window $\sim \Gamma _{\mathbf{k}}$
around $\epsilon _{\mathbf{k}}$. Then, it is easy to see that $\mathcal{A}\left(
\mathbf{k},\omega \right) $ is an even function and $d\mathcal{A}\left(
\mathbf{k,}\omega \right)/d\omega $ is an odd function centered around $\epsilon _{%
\mathbf{k}}$. Hence, after the integration, $\alpha $ approaches zero. In
the critical regime, there is no well-defined quasi-particles, and the
spectral function usually behaves as \cite{Sachdev2011}
\begin{equation}
\mathcal{A}(\mathbf{k},\omega )\sim \frac{\Theta \left( \omega -\varepsilon _{\mathbf{k%
}}\right) }{\left( \omega -\varepsilon _{\mathbf{k}}\right) ^{\eta }}\,,
\end{equation}%
where $\eta $ is a critical exponent. In the high temperature limit, we have
approximated $f_{\mathrm{B}}(\omega )\simeq e^{-\beta \left( \omega -\mu
\right) }$ in integration. Thus we have,
\begin{equation*}
\alpha \sim 4\pi \bar{n}\eta \int_{\varepsilon _{\mathbf{k}}}^{\infty
}d\omega \frac{e^{-\beta \left( \omega -\mu \right) }}{\left( \omega
-\varepsilon _{\mathbf{k}}\right) ^{2\eta +1}}\sim 4\pi \bar{n}\eta \frac{%
e^{-\beta (\varepsilon _{\mathbf{k}}-\mu )}}{T^{2\eta }}.
\end{equation*}

\section{The Validity of the Wick's Contraction\label{Wick}}

As mentioned above, the Wick's expansion ignores the vertex corrections.
Hence, our following discussions will focus on vertex corrections. The first
order perturbation contribution to $\mathsf{G}(i\omega _{n})$ is given by
\begin{eqnarray}
&&\mathsf{G}^{\left( 1\right) }(i\omega _{n})=8U\bar{n}\sum\limits_{\mathbf{k%
}^{\prime },mm^{\prime }}g_{0}(\mathbf{k}^{\prime },i\nu _{m^{\prime
}})g_{0}(\mathbf{k}^{\prime },i\nu _{m^{\prime }}-i\omega _{n})  \notag \\
&&\times g_{0}(\mathbf{k},i\nu _{m})g_{0}(\mathbf{k},i\nu _{m}-i\omega _{n})
\notag \\
&&+4U\sum\limits_{\mathbf{k}^{\prime }\mathbf{q},mm^{\prime }\ell }g_{0}(%
\mathbf{k}^{\prime }-\mathbf{q},i\nu _{m^{\prime }}-i\nu _{\ell })  \notag \\
&&\times g_{0}(\mathbf{k}^{\prime },i\nu _{m^{\prime }})g_{0}(\mathbf{k}%
,i\nu _{m})g_{0}(\mathbf{k}+\mathbf{q},i\nu _{m}+i\nu _{\ell })  \notag \\
&&\times \left[ g_{0}(\mathbf{k},i\nu _{m}-i\omega _{n})+g_{0}(\mathbf{k}%
,i\nu _{m}+i\omega _{n})\right]
\end{eqnarray}%
where $g_{0}(\mathbf{k},i\nu _{m})$ is the free two-point Green's function. This
equation can be rewritten into%
\begin{eqnarray}
&&\mathsf{G}^{\left( 1\right) }(i\omega _{n})=8U\bar{n}\Pi _{0}\left(
0,i\omega _{n}\right)   \notag \\
&&\times \sum\limits_{m}g_{0}(\mathbf{k},i\nu _{m})g_{0}(\mathbf{k},i\nu
_{m}-i\omega _{n})  \notag \\
&&+4U\sum\limits_{\mathbf{q},m\ell }g_{0}(\mathbf{k}+\mathbf{q},i\nu
_{m}+i\nu _{\ell })  \times \Pi _{0}\left( \mathbf{q},i\nu _{\ell }\right) g_{0}(\mathbf{k}%
,i\nu _{m})  \notag \\
&&\times \left[ g_{0}(\mathbf{k},i\nu _{m}-i\omega _{n})+g_{0}(\mathbf{k}%
,i\nu _{m}+i\omega _{n})\right],
\end{eqnarray}
where 
\begin{equation}
\Pi _{0}(\mathbf{q},i\nu _{\ell })=\sum\limits_{\mathbf{q},m^{\prime
}}g_{0}(\mathbf{k}^{\prime },i\nu _{m^{\prime }})g_{0}(\mathbf{k}^{\prime }-%
\mathbf{q},i\nu _{m^{\prime }}-i\nu _{\ell })
\end{equation}
and $\Pi _{0}(\mathbf{q},i\nu _{\ell })$ is the free density
fluctuation. By resuming the high-order diagrams, a significant part of 
contributions can be obtained by replacing the free Green's functions $g_{0}$
and $\Pi _{0}$ with the full Green's functions $g$ and $\Pi $ respectively.
Then, we have
\begin{eqnarray}
&&\mathsf{G}(i\omega _{n})=\mathsf{G}^{W}(i\omega _{n})  \notag \\
&&+8U\bar{n}\Pi \left( 0,i\omega _{n}\right) \sum\limits_{m}g(\mathbf{k}%
,i\nu _{m})g(\mathbf{k},i\nu _{m}-i\omega _{n})  \notag \\
&&+4U\sum\limits_{\mathbf{q},m\ell }g(\mathbf{k}+\mathbf{q},i\nu _{m}+i\nu
_{\ell })\Pi \left( \mathbf{q},i\nu _{\ell }\right) g(\mathbf{k},i\nu
_{m})  \notag \\
&&\times \left[ g(\mathbf{k},i\nu _{m}-i\omega _{n})+g(\mathbf{k},i\nu
_{m}+i\omega _{n})\right],
\end{eqnarray}%
where $\mathsf{G}^{W}$ is the part given by the Wick's contraction defined in Eq. \ref{eq:G_tau}
We can see that the contribution of the vertex corrections are controlled by 
the density fluctuations.

We argue that Wick's contraction is a reasonable approximation for two reasons \cite{Pan}. The vertex corrections can be safely ignored in the weakly interacting superfluid phase because the interaction strength is weak. In the strongly interacting regime, the system is either a Mott insulator or a critical regime. In the Mott insulator, the density fluctuation is gapped. In the critical regime, the compressibility continuously approaches zero. Since the vertex corrections are controlled by the density fluctuations, the contributions of vertex corrections are also highly suppressed.

\end{document}